\documentclass{vldb}

\usepackage{booktabs} 
\usepackage{amsmath}
\usepackage{graphicx,xspace,verbatim,comment}
\usepackage{hyperref,array,color,balance,multirow}
\usepackage{balance,float,url,amsfonts,alltt}
\usepackage{mathtools,rotating,amsmath,amssymb}
\usepackage{color,ifpdf,fancyvrb,array}
\usepackage{etoolbox,listings,subcaption}
\usepackage{bigstrut,morefloats}
\usepackage[boxruled]{algorithm2e}
\usepackage{pbox}

\newcommand{\eat}[1]{}

\newenvironment{packeditems}{
\begin{itemize}
  \setlength{\itemsep}{1pt}
  \setlength{\parskip}{0pt}
  \setlength{\parsep}{0pt}
}{\end{itemize}}

\title{Are Key-Foreign Key Joins Safe to Avoid \\when Learning High-Capacity Classifiers?\\}

\numberofauthors{1}
\author{
\alignauthor Vraj Shah$^1$\hspace{20mm}Arun Kumar$^1$\hspace{20mm}Xiaojin Zhu$^2$\\
\affaddr{\vspace{2mm}$^1$University of California, San Diego\hspace{20mm}$^2$University of Wisconsin-Madison}
\email{\{vps002, arunkk\}@eng.ucsd.edu, jerryzhu@cs.wisc.edu}
}

\begin{document}

\maketitle

\begin{abstract}
Machine learning (ML) over relational data is a booming area of the database industry and 
academia. While several projects aim to build scalable and fast ML systems, little work has 
addressed the pains of \textit{sourcing} data and features for ML tasks. 
Real-world relational databases typically have many tables (often, dozens) and data scientists 
often struggle to even obtain and join all possible tables that provide features for ML. 
In this context, Kumar et al.~showed recently that key-foreign key dependencies (KFKDs) 
between tables often lets us avoid such joins without significantly affecting prediction 
accuracy--an idea they called ``avoiding joins safely." While initially controversial, this idea 
has since been used by multiple companies to reduce the burden of data sourcing for ML.
But their work applied only to linear classifiers. In this work, we verify if their results hold 
for three popular high-capacity classifiers: decision trees, non-linear SVMs, and ANNs. We conduct an 
extensive experimental study using both real-world datasets and simulations to analyze the 
effects of avoiding KFK joins on such models. Our results show that these high-capacity classifiers 
are surprisingly and counter-intuitively \textit{more} robust to avoiding KFK joins compared to 
linear classifiers, refuting an intuition from the prior work's analysis. We explain this behavior 
intuitively and identify open questions at the intersection of data management and ML 
theoretical research. All of our code and datasets are available for download from 
\url{http://cseweb.ucsd.edu/~arunkk/hamlet}.
\end{abstract}

\section{Introduction}
The data management community has long studied how to integrate ML with data systems 
(e.g.,~\cite{madlib,bismarck,riot}), how to scale ML (e.g.,~\cite{systemml,graphlab}), and 
how to use database ideas to improve ML tasks (e.g.,~\cite{columbus2,mlbase}). 
However, little work has tackled the pains of \textit{sourcing} data for ML tasks in the first place, 
especially, how fundamental data properties affect end-to-end data workflows for ML tasks~\cite{brainwash}.
In particular, real-world relational databases often have many tables connected by database 
dependencies such as \textit{key-foreign key dependencies} (KFKDs)~\cite{cowbook}. Thus, given 
an ML task, data scientists almost always \textit{join} multiple tables because they like to obtain 
more \textit{features} for ML models~\cite{orion}. But from conversations with data scientists 
at many enterprise and Web companies, we learned that even this simple process of procuring 
features by joining tables could be painful in practice because different tables could be 
``owned" by different teams with different access restrictions. This slows down the ML
analytics lifecycle. Furthermore, recent reports of Google's production ML systems show
that features that yield marginal benefits incur high ``technical debt" that decreases code 
mangeability and increases costs~\cite{techdebt,googletutorial}.

In this context, Kumar et al.~\cite{hamlet} showed that one can often omit entire tables by 
exploiting KFKDs in the database \textit{schema}. That is, one can ignore a table 
without even looking at its contents (i.e., ``avoid the join"), but crucially, do so without 
significantly affecting ML accuracy (i.e., ``safely"). 
The basis for this dramatic capability is that a KFK join creates a \textit{functional dependency} 
(FD) between a \textit{foreign key feature} and the \textit{foreign features} brought in by the 
join.\footnote{While KFKDs are not the same as FDs~\cite{avibook}, assuming features have
``closed" domains, they behave essentially as FDs in the output of the join~\cite{hamlet}.}

\vspace{1mm}
\noindent{\textbf{Example (based on~\cite{hamlet}).}}
Consider a common classification task: predicting \textit{customer 
churn}. The data scientist starts with the main table for training (simplified for exposition): 
\texttt{Customers} (\underline{\texttt{CustomerID}}, \texttt{Churn}, \texttt{Gender}, \texttt{Age}, 
\texttt{Employer}). \texttt{Churn} is the target, while \texttt{Gender}, \texttt{Age}, and
\texttt{Employer} are features. So far, this is a standard classification task. 
She then notices the table \texttt{Employers} (\underline{\texttt{Employer}}, \texttt{State}, 
\texttt{Revenue}) in her database with extra features about customers' employers.
\texttt{Customers}.\texttt{Employer} is thus a foreign key feature connecting these tables. 
She joins the tables to bring in the foreign features (about employers) because she has a hunch that 
customers employed by rich companies in coastal states might be less likely to churn. She then tries 
different classifiers, e.g., logistic regression or decision trees.

The analysis in~\cite{hamlet} revealed a dichotomy in how safe it is to avoid a join from an accuracy
standpoint: in terms of the bias-variance trade-off, avoiding a join is unlikely to increase bias but 
it might significantly increase variance, since foreign key features often have larger domains than 
foreign features. \textit{In simple terms, avoiding joins might cause extra overfitting}. 
But this extra overfitting subsides with more training examples. 
In~\cite{hamlet}, the \textit{tuple ratio} quantifies this behavior; in our example, it is the ratio of the 
number of labeled customers to the number of employers. When the tuple ratio is above a certain VC 
dimension-based threshold, we can safely avoid the join. 
For simpler classifiers with linear VC dimensions (e.g., logistic regression and Naive 
Bayes), this threshold was about $20$. Since there were public real-world datasets that satisfied this 
threshold, this idea of avoiding joins safely could be empirically validated. 

While initially controversial, the idea of avoiding joins safely has been adopted by many data scientists, 
including at Facebook, LogicBlox, and MakeMyTrip~\cite{perscomm}. Based on the value of the 
easy-to-compute tuple ratio, which only needs the foreign table's cardinality rather than the table itself, 
the data scientist can decide if they want to avoid the join or procure the extra table. However, the 
results in~\cite{hamlet} had a major caveat--they applied only to linear classifiers. In fact, their VC 
dimension-based analysis suggested that the tuple ratio thresholds might be too high for high-capacity 
non-linear classifiers, potentially rendering this idea of avoiding joins safely inapplicable to such
classifiers in practice.

\textit{In this paper, we perform a comprehensive empirical and simulation study and analysis to 
verify (or refute) the applicability of the idea of avoiding joins safely to three popular high-capacity 
classifiers: decision trees, SVMs, and ANNs.}

The natural expectation is that these complex models, some with infinite VC dimensions, will likely face 
larger extra overfitting by avoiding joins compared to linear classifiers. Surprisingly, our results 
show that their behavior is the \textit{exact opposite}! We start by rerunning the experiments on the 
real-world datasets with KFK joins from~\cite{hamlet} for these models.\footnote{But for simplicity 
and ease of comparison of all the models, we binarize all classification tasks.} Irrespective of whether 
we use linear classifiers or the higher-capacity classifiers, the same set of joins turn out to be safe to avoid.
Furthermore, on the datasets that had joins that were not safe to avoid, the decrease in accuracy caused 
by avoiding said joins (unsafely) was lower for the higher-capacity classifiers compared to the linear classifiers.
In other words, \textit{our work refutes an intuition from the VC dimension-based analysis of~\cite{hamlet} 
and shows that these popular high-capacity classifiers are counter-intuitively \textit{comparably} or 
\textit{more} robust to avoiding joins than linear classifiers, not less}.

To understand the above surprising behavior in depth, we conduct an in-depth Monte Carlo-style simulation study 
to ``stress test" how safe it is to avoid the join. We use decision trees for the simulation study, since they were 
the most robust to avoiding joins. We generate data for a two-table KFK join and embed various ``true'' distributions 
for the target. This includes a known ``worst-case'' scenario for avoiding joins with linear 
classifiers (i.e., the holdout test errors blow up)~\cite{hamlet}. We vary different properties of the data and the true distribution: 
numbers of features in each base table, numbers of training examples, foreign key domain size, noise in the data, and foreign key skew. 
In very few of these cases does avoiding the join cause the error to rise beyond $1\%$! Indeed, the only scenario where avoiding the join 
caused significantly higher overfitting was when the tuple ratio was less than $3$; this scenario arose in only $1$ of the $7$ real datasets.
These results are in stark constrast to the results for linear classifiers.

Our counter-intuitive results raise new research questions at the intersection of data management and ML theory.
In particular, there is a need to formalize the effects of KFKDs/FDs on the behavior of decision trees, SVMs, and ANNs.
As a step in this direction, we analyze and intuitively explain the behavior of decision trees and SVMs.
\eat{
Essentially, we explain why RBF-SVMs with foreign key features behave somewhat like a 1-nearest neighbor classifier due to 
the high dimensionality of foreign key features with one-hot encoding. 
While this leads to a form of \textit{memorization} of the foreign key's domain, this seems to have little effect on the 
model's generalization or test errors. This is similar to how memorization seems to occur in deep neural networks~\cite{rechtdnn}, 
but a key difference in our setting is that such memorization does not apply to \textit{all} features. 
We also discuss why decision trees are robust to operating with foreign key features. 
}
Other open questions include the implications of more general database dependencies such as embedded multi-valued dependencies on 
the behavior of such models and the implications of all database dependencies for other ML tasks such as regression and clustering.
We believe that solving these fundamental questions could lead new analytics systems functionalities that make it easier to use ML
for data analytics.

\eat{
We extend the worst-case simulation scenario for linear models by replicating the foreign feature that determines the target multiple times. The idea is to make a model that uses 
the foreign key feature alone to overfit more than one that uses the foreign features. In particular, for the RBF-SVM, this scenario demonstrates that it behaves 
more similarly to a 1-nearest neighbor classifier when using the foreign key feature but less so when the number of relevant foreign features are increased.
}

Finally, we observed that foreign key features cause two new practical bottlenecks for data scientists, especially with decision trees. 
First, the sheer size of their domains makes it hard to interpret and visualize the trees. 
Second, some foreign key values may not have any training examples even if they are known to be in the domain. 
We identify and adapt standard heuristics from the literature to resolve these bottlenecks and verify their effectiveness empirically.
\eat{ 
For the first bottleneck, we propose simple lossy 
\textit{domain compression} methods that are configurable with a user-given size budget. For the second bottleneck, we propose a form of foreign key \textit{smoothing} 
that exploits foreign features as side information. We validate the accuracy of these techniques using both real and synthetic datasets.
}

\vspace{1mm}
\noindent Overall, the contributions of this paper are as follows:

\begin{packeditems}
\item To the best of our knowledge, this is the first paper to analyze the effects of avoiding KFK joins on three popular 
high-capacity classifiers: decision trees, SVMs, and ANNs.
We present a comprehensive empirical study that refutes an intuition from prior work and shows that these classifiers 
are counter-intuitively more robust to avoiding joins than linear classifiers.

\item We conduct an in-depth simulation study with a decision tree to assess the effects of various data properties on
how safe it is to avoid a KFK join.

\item We present an intuitive analysis to explain the behavior of decision trees and SVMs when joins are avoided. We identify 
open questions for research at the intersection of data management and ML theory.

\item We identify two practical bottlenecks with foreign key features and verify the effectiveness of standard heuristics in resolving them.
\end{packeditems}

\vspace{-3mm}
\paragraph*{\textbf{Outline}} Section 2 presents our notation, assumptions, and scope. 
Section 3 presents results on the real data, while Section 4 presents our simulation study. Section 5 presents 
our intuitive analysis of the results and identifies open research questions.
Section 6 verifies the techniques to make foreign key features more practical. 
We discuss related prior work in Section 7 and conclude in Section 8.

\section{Preliminaries}

\subsection{Notation}
The setting we focus on is the following: the dataset has a set of tables in the \textit{star schema} with KFK dependencies (KFKDs).
Star schemas are ubiquitous in many applications, including retail, insurance, Web security, and recommendation systems~\cite{cowbook,hamlet,orion}.
The fact table, which has the target variable, is denoted \textbf{S}. It has the schema $\textbf{S}(\underline{SID},Y, \textbf{X}_S, FK_1, \dots, FK_q)$.
A dimension table is denoted $\textbf{R}_i$ ($i = 1$ to $q$) and it has the schema $\textbf{R}_i(\underline{RID_i},\textbf{X}_{R_i})$.
$Y$ is the target variable (class label), $\textbf{X}_S$ and $\textbf{X}_{R_i}$ are vectors (sequences) of features, $RID_i$ is the primary key
of $\textbf{R}_i$, while $FK_i$ is a foreign key feature that refers to $\textbf{R}_i$. 
We call $\textbf{X}_S$ \textit{home} features and $\textbf{X}_{R_i}$ \textit{foreign} features.
For ease of exposition, we also treat \textbf{X} as a \textit{set} of features since the order among features is immaterial in our setting.
Let \textbf{T} denote the output of the projected equi-join (key-foreign key, or KFK for short) query that constructs the full training dataset by 
\textit{concatenating} the features from all base tables: $\textbf{T} \leftarrow \pi(\textbf{R} \bowtie_{RID=FK} \textbf{S})$. In general, its schema is 
$\textbf{T}(\underline{SID},Y,\textbf{X}_S,FK_1,\dots,FK_q,\textbf{X}_{R_1},\dots,\textbf{X}_{R_q})$. In contrast to our setting, traditional ML 
formulations do not distinguish between home features, foreign keys, and foreign features. The number of tuples in \textbf{S} (resp.~\textbf{R}) is 
denoted $n_S$ (resp.~$n_R$); the number of features in $\textbf{X}_S$ (resp.~$\textbf{X}_R$) is denoted $d_S$ (resp.~$d_R$). Without loss of generality, we
assume that the join is not selective, which means $n_S$ is also the number of tuples in \textbf{T}. $\mathcal{D}_{FK}$ denotes the domain of $FK$ 
and by definition, $|\mathcal{D}_{FK}| = n_R$. We call $\frac{n_S}{n_R}$ the \textit{tuple ratio}.
Note that our setting is different from the statistical relational learning (SRL) setting, which deals with joins that violate the IID 
assumption and duplicate labeled examples from \textbf{S}~\cite{srlbook}. KFK joins do not cause such duplication and thus, regular IID models
are typically used in this setting.

\eat{
\paragraph*{Example}
Consider a common classification task for which ML classifiers are used, predicting \textit{customer churn}.
The fact table is \texttt{Customers} (\underline{\texttt{CustomerID}}, \texttt{Churn}, \texttt{Gender}, \texttt{Age}, \texttt{Employer}, \texttt{ZipCode}).
\texttt{Employer} and \texttt{Zipcode} are foreign key features that refer respectively to a customer's employer (e.g., Google or Microsoft) and the area 
where a customer lives. The respective dimension tables are \texttt{Employers} (\underline{\texttt{Employer}}, \texttt{State}, \texttt{Revenue}) 
and \texttt{Areas} (\underline{ZipCode}, \texttt{CrimeRate}, \texttt{AccidentRate}).
In such scenarios, data scientists typically join all base tables to bring in the extra features from the dimension tables. In this case, they might do so 
because of a hunch that customers employed by rich corporations in coastal states and living in ``safe'' areas are unlikely to churn.
}

\subsection{Assumptions and Scope}
For the sake of tractability, in this paper, we adopt some assumptions from~\cite{hamlet}. In particular, we assume the features are 
\textit{categorical}. Numeric features can be discretized using standard techniques such as binning~\cite{mitchellbook}. 
We also focus on binary classification but our ideas can be easily applied to multi-class targets as well. 
We assume that the foreign key features ($FK_i$) are not (primary) keys in the fact table, e.g., \texttt{Employer} does 
not uniquely identify a customer.\footnote{Primary keys in the fact table are \textit{not} generalizable features, unlike foreign keys.}
Finally, we also do not study the ``cold start'' issue because it is orthogonal to the focus of this paper~\cite{coldstart}. Thus, all features 
have known finite domains, possibly including a special ``Others'' placeholder to temporarily handle hitherto unseen values. 
In our example, this means that both \texttt{Employer} and \texttt{Gender} have known finite domains.
In general, $FK_i$ can take values only from the given set of $\textbf{R}_i.RID_i$ values (new $FK_i$ values 
are mapped to ``Others''). Since ML models are rebuilt periodically in practice, new information can then be added to expand feature domains. 
We emphasize that our goal is \textit{not} to create new ML or feature selection algorithms, nor is to compare which algorithms yield 
the best accuracy or runtime. Our goal is to expose and analyze how KFKDs/FDs enable us to dramatically discard foreign features a priori 
when learning some popular high-capacity classifiers.

\begin{table}[t]
\centering
\includegraphics[width=0.99\linewidth]{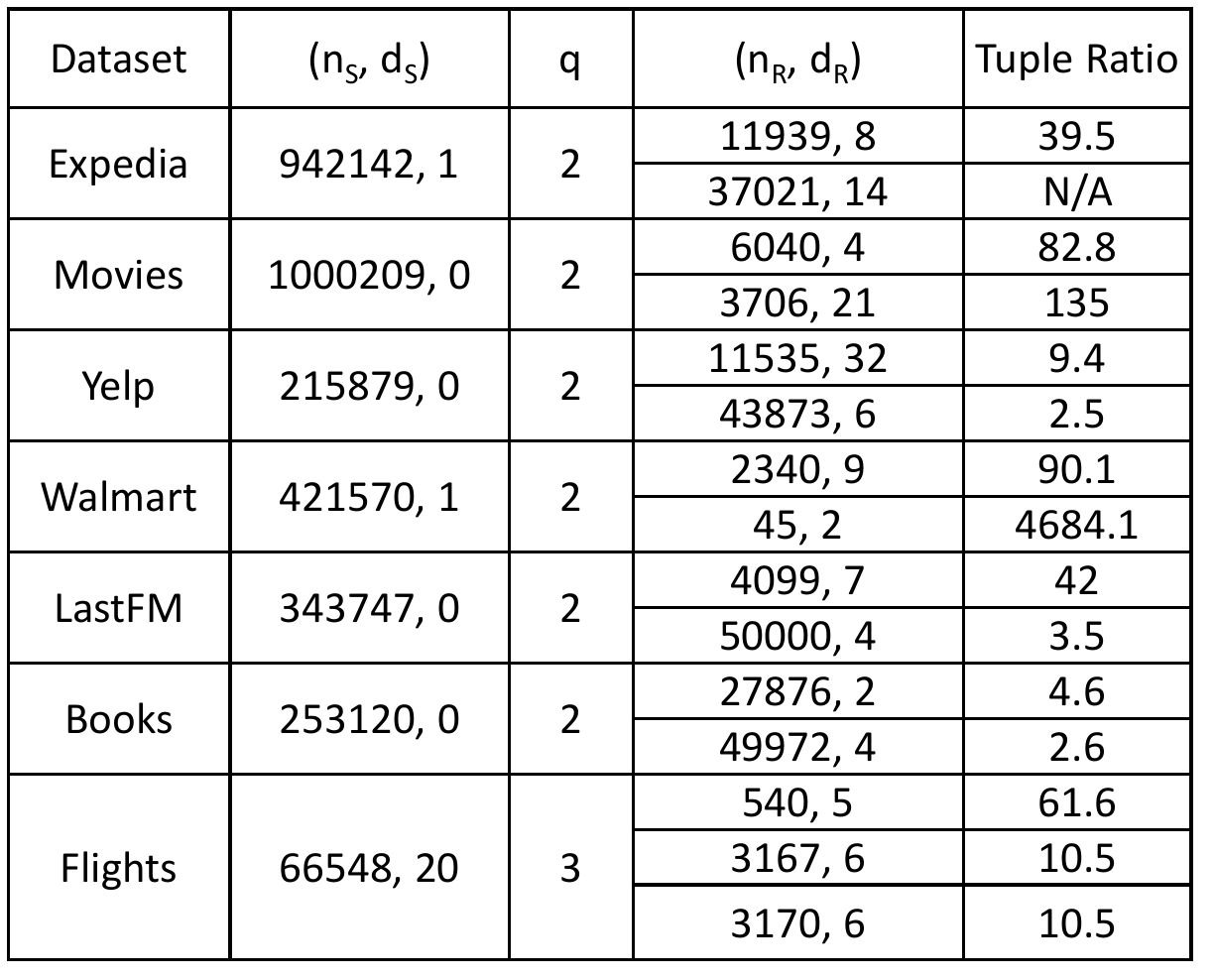}
\caption{Dataset statistics. $q$ is the number of dimension tables. All features are categorical. $n_S$ is the number of labeled examples, also 
overloaded to mean the number of training examples ($50\%$ as many). So, the tuple ratio listed is actually $50\% \times n_S / n_R$. N/A means the 
corresponding dimension table can never be discarded because its corresponding foreign key has an ``open'' domain and can never be used as a feature.
}
\label{Table:datastats}
\end{table}

\section{Empirical Study with Real Data}

We now present our detailed empirical study using real-world datasets on 10 classifiers, including 7 high-capacity classifiers (CART decision tree with gini, 
information gain, and gain ratio; SVM with RBF and quadratic kernels; multi-layer perceptron ANN; the ``braindead" 1-nearest neighbor), and 3 linear classifiers 
(Naive Bayes with backward selection, logistic regression with L1 regularization, and linear SVM).
We also conducted experiments with a few other feature selection techniques for the linear classifiers: Naive Bayes with forward selection and filter methods, 
as well as logistic regression L2 regularization Since these additional linear classifiers did not provide any new insights, we omit them due to space constraints.

\subsection{Datasets}
We take the seven real datasets from~\cite{hamlet}; these are originally from Kaggle, GroupLens, \url{openflights.org}, \url{mtg.upf.edu/node/1671}, and \url{last.fm}.
Two datasets have binary targets (Flights and Expedia); the others have multi-class ordinal targets. For the sake of simplicity, we binarize all targets 
for this paper by grouping ordinal targets into lower and upper halves (this change does not affect our overall conclusions). 
The dataset statistics are provided in Table~\ref{Table:datastats}. 
We briefly describe the task for each dataset and explain what the foreign features are. More details about their schemas, including the list 
of all features are already in the public domain (listed in~\cite{hamlet}). 
\textit{All of our datasets, scripts, and code are available for download on our project webpage\footnote{\url{http://cseweb.ucsd.edu/~arunkk/hamlet}} 
to make reproducibility easier.}

\begin{table*}[t]
\centering
\includegraphics[width=0.99\linewidth]{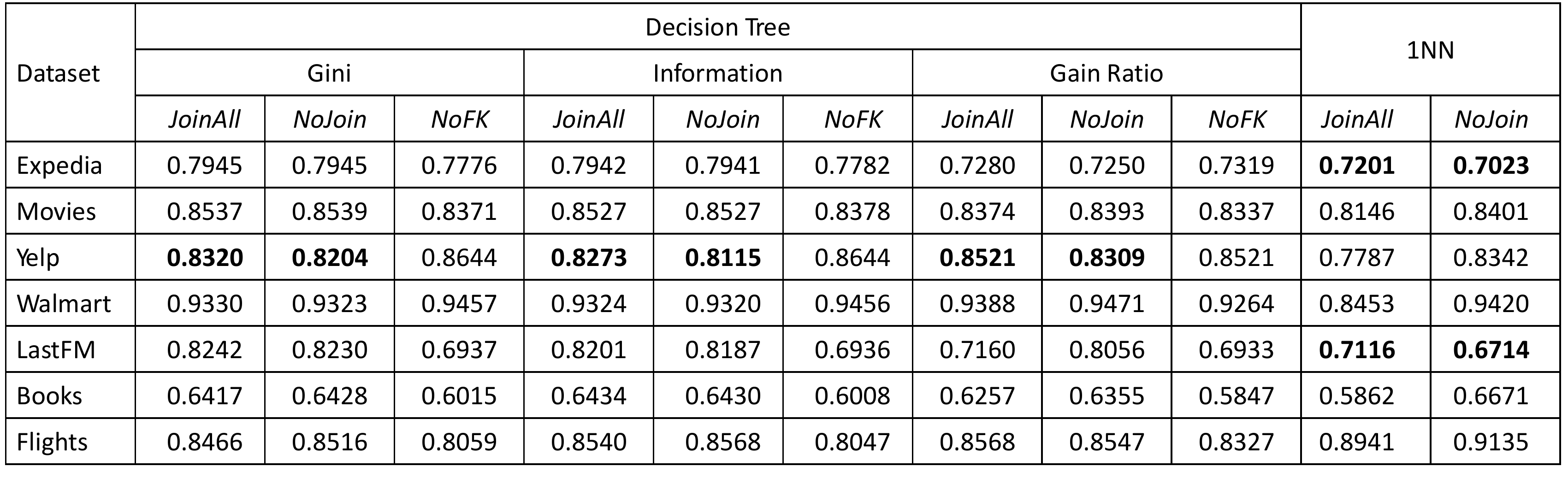}
\caption{Holdout test accuracy on the real-world datasets for the three decision trees and 1-NN. 
Our goal is \textit{not} to compare the accuracy across ML models, but rather compare
the accuracy of \textit{JoinAll} and \textit{NoJoin} within each model.
The bold font marks the only cases where the accuracy of \textit{NoJoin} is at least 1\% lower 
than the accuracy of \textit{JoinAll}.
}
\label{Table:RealTest1}
\end{table*}

\begin{table*}[t]
\centering
\includegraphics[width=0.99\linewidth]{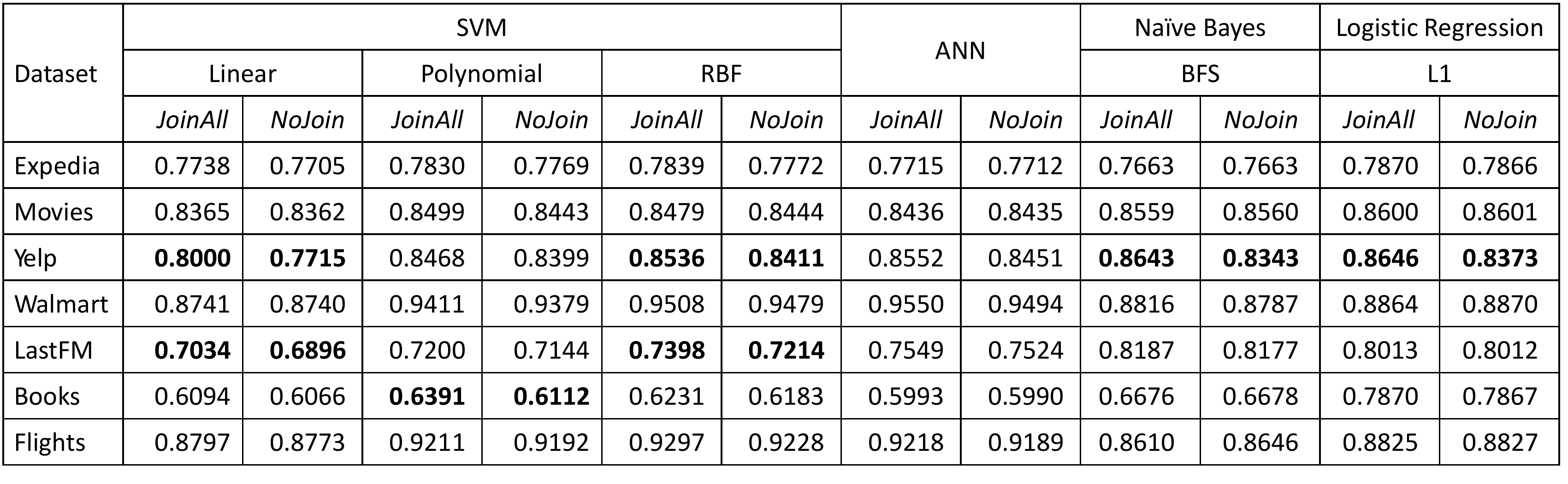}
\caption{Holdout test accuracy on the real-world datasets for the three SVMs, ANN, Naive Bayes, and logistic regression.
Our goal is \textit{not} to compare the accuracy across ML models, but rather compare
the accuracy of \textit{JoinAll} and \textit{NoJoin} within each model.
The bold font marks the only cases where the accuracy of \textit{NoJoin} is at least 1\% lower 
than \textit{JoinAll}.
}
\label{Table:RealTest2}
\end{table*}

\vspace{1mm}
\textit{Walmart}: predict if department-wise sales will be high using past sales (fact table) joined with stores and weather/economic indicators (two dimension tables).

\vspace{1mm}
\textit{Flights}: predict if a route is codeshared by using other routes (fact table) joined with airlines, source, and destination airports (three dimension tables).

\vspace{1mm}
\textit{Yelp}: predict if a business will be rated highly using past ratings (fact table) joined with users and businesses (two dimension tables).

\vspace{1mm}
\textit{MovieLens}: predict if a movie will be rated highly using past ratings (fact table) joined with users and movies (two dimension tables).

\vspace{1mm}
\textit{Expedia}: predict if a hotel will be ranked highly using past search listings (fact table) joined with hotels and search events (two dimension tables 
but one foreign key, viz., the search ID, has an ``open'' domain, i.e., past values will not be seen in the future, which makes it unusable as a feature).

\vspace{1mm}
\textit{LastFM}: predict if a song will be played often using past play level information (fact table) joined with users and artists (two dimension tables).

\vspace{1mm}
\textit{Books}: predict if a book will be rated highly using past ratings (fact table) joined with readers and books (two dimension tables).

\subsection{Methodology}
Each dataset is pre-split 50\%:25\%:25\% for training, validation (during feature selection and hyper-parameter tuning), and holdout testing. We retain the splits as is.
We compare two approaches: \textit{JoinAll}, which joins all base tables to provide all features to the classifier (the current widespread practice), and \textit{NoJoin},
which avoids all foreign features a priori (the approach we study). We compare these two approaches for all the 10 classifiers mentioned earlier. For additional insights, 
we also include a third approach for the decision trees: \textit{NoFK}, which is essentially the same as  \textit{JoinAll} but with all foreign key features dropped a priori.
We used the popular R packages ``rpart'' for the decision trees\footnote{Except for the gain ratio case for which we used the ``CORElearn'' package in R.} 
and ``e1071'' for the SVMs.  For the ANNs, we used the popular Python library Keras on top of TensorFlow.
For Naive Bayes, we used the code from~\cite{hamlet}, while for logistic regression with L1 regularization, we used the popular R package ``glmnet."
We use the validation set for hyper-parameter tuning using a standard grid search for each classifier with the grids described in detail below. Note that 
for Naive Bayes, there is no hyper-parameter tuning.

\vspace{2mm}
\textit{Decision Trees}: There are two hyper-parameters to tune: \textit{minsplit} and \textit{cp}. \textit{minsplit} is the number of observations that 
must exist in a node for a split to be attempted. Any split that does not improve the fit by a factor of \textit{cp} is pruned off. 
The grid is set as follows: \textit{minsplit} $\in \{ 1, 10, 100, 10^3\}$ and \textit{cp} $\in \{10^{-4}, 10^{-3}, 0.01, 0.1, 0\}$   

\vspace{2mm}
\textit{RBF-SVM}: There are two hyper-parameters to tune: \textit{C} and $\gamma$. 
\textit{C} controls the cost of misclassification. $\gamma > 0$ controls the bandwidth in the Gaussian kernel 
(given two data points $x_i$ and $x_j$): $k(x_i,x_j) = \exp(-\gamma \cdot \lVert{x_i - x_j} \rVert ^2 )$.
The grid is set as follows: \textit{C} $\in \{10^{-1}, 1, 10, 100, 10^3\}$ and $\gamma \in \{10^{-4}, 10^{-3}, 0.01, 0.1, 1, 10\}$.\footnote{On 
\textit{Movies} and \textit{Expedia} alone, we perform an extra fine tuning step with 
$\gamma \in \{2^{-7}, 2^{-6},2^{-5}, 2^{-4}, 2^{-3}, 2^{-2},2^{-1}, 1,2, 2^{2}, 2^{3}\}$ to improve accuracy.}

\vspace{2mm}
\textit{Quadratic-SVM}: We tune the same hyper-parameters \textit{C} and $\gamma$ for the polynomial kernel of degree 2:
$k(x_i,x_j) = (-\gamma \  x_i ^T \cdot x_j)^{degree}$. We use the same grid as RBF-SVM.
\vspace{2mm}

\textit{Linear-SVM}: We tune the \textit{C} hyper-parameter for the linear kernel: 
$k(x_i,x_j) = x_i ^T \cdot x_j $, \textit{C} $\in \{10^{-1}, 1, 10, 100, 10^3\}$.

\vspace{2mm}
\textit{ANN}: The multi-layer perceptron architecture comprises of 2 hidden units with 256 and 64 neurons respectively. 
Rectified linear unit (ReLU) is used as the activation function. In order to allow penalties on layer parameters, we do $L_2$ regularization, with 
the regularization parameter tuned using the following grid axis: $\{ 10^{-4}, 10^{-3}, 10^{-2} \}$. We choose the popular Adam stochastic 
gradient optimization algorithm~\cite{adam} with the learning rate tuned using the following grid axis:  $\{ 10^{-3}, 10^{-2}, 10^{-1} \}$. 
The other hyper-parameters of the Adam algorithm used the default values.

\vspace{2mm}
\textit{Logistic Regression}: The glmnet package performs automatic hyper-parameter tuning for the L1 regularizer, as well as the optimization algorithm.
However, it has three parameters to specify a desired convergence threshold and a limit on the execution time: \textit{nlambda}, which we set to 100,
\textit{maxit}, which we set to 10000, and \textit{thresh}, which we set to 0.001.

Tables~\ref{Table:RealTest1} and~\ref{Table:RealTest2} present the holdout test accuracy results for all the models on all the datasets.

\begin{table}[t]
\centering
\includegraphics[width=\columnwidth,height=\textheight,keepaspectratio]{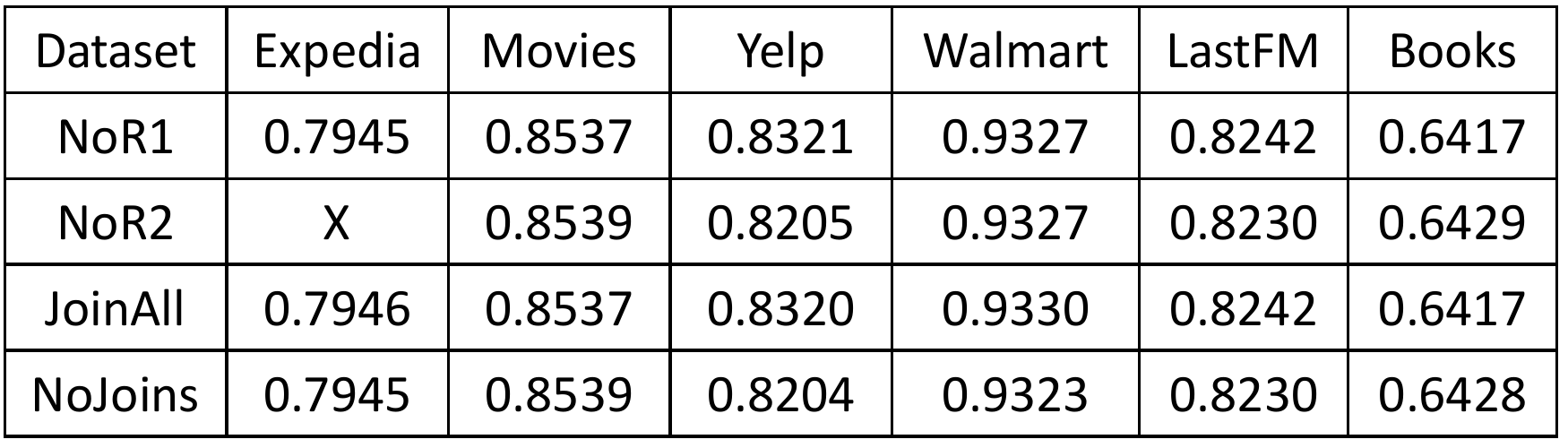}
\begin{align*}
\textbf{Flights} : NoR1: 0.8466 \quad NoR2: 0.8490 \quad NoR3: 0.8483 \\
No R1,R2: 0.8488 \quad No R1,R3: 0.8481 \quad No R2,R3: 0.8519
\end{align*}
\caption{Robustness study for discarding dimension tables on the real-world datasets with a Gini decision tree.}
\label{Table:robustness}
\end{table}

\subsection{Results}

\vspace{-1mm}
\subsubsection*{Accuracy}
Our first and most important observation is that for almost all the datasets (\textit{Yelp} being the exception) and for all three split criteria, the accuracy of the 
decision tree is comparable (a gap of within 1\%) between \textit{JoinAll} and \textit{NoJoin}.\footnote{Except for gain ratio on \textit{LastFM}, where \textit{NoJoin}
is actually significantly \textit{more} accurate than \textit{JoinAll}!} The trend is virtually the same for the RBF-SVM and ANN as well. 
We also observe that the trend is virtually the same for the linear models as well! Thus, regardless of whether our classifier is linear or higher capacity, the relative 
behavior of \textit{NoJoin} vis-a-vis \textit{JoinAll} is virtually the same. 
These results represent our key counter-intuitive finding: joins are no less safe to avoid with the high-capacity classifiers than with the 
linear classifiers. The absolute accuracy of the high-capacity classifiers is often significantly higher than the linear classifiers, which is as expected but is 
also orthogonal and irrelevant to this paper's focus.
Interestingly, on the \textit{Yelp} dataset, in which both joins are known to be \textit{not} safe to avoid with the linear classifiers~\cite{hamlet}, \textit{NoJoin} 
correctly sees a large reduction in accuracy from \textit{JoinAll}--about $0.03$. 
However, the drop in accuracy is smaller for the high-capacity classifiers, e.g., the RBF-SVM, Gini decision tree, and ANN all see a drop of only about $0.01$!
This suggests that these high-capacity classifiers are sometimes counter-intuitively \textit{more} robust than linear classifiers to avoiding joins.

As for \textit{NoFK}, it often has much lower accuracy than both \textit{JoinAll} and \textit{NoJoin} on all three forms of decision trees. For example, on 
\textit{Flights}, the drop is about $0.05$. This reaffirms the importance of foreign key features; as such, it is known that dropping foreign key features could 
cause the bias to shoot up with linear classifiers~\cite{hamlet}. A similar scenario arises for the high-capacity classifiers too.
Interestingly, in some cases (e.g., Gini on \textit{Flights} and gain ratio on \textit{Books}), \textit{NoJoin} has marginally higher accuracy than \textit{JoinAll}.

To understand the above  results more deeply, we conduct a ``robustness'' experiment by discarding dimension tables one at a time instead of all together. 
Table~\ref{Table:robustness} presents this experiment's results for the Gini decision tree.
Discarding each dimension table one at a time (and also two at a time in the case of \textit{Flights}) did not differ much from \textit{NoJoin}, except for \textit{Yelp}.
On \textit{Yelp}, the accuracy drops only when $\textbf{R}_2$ (users table) is dropped. 
From Table~\ref{Table:datastats}, we find that the tuple ratio for $\textbf{R}_2$ in \textit{Yelp} is extremely low: $2.5$. 
That is, there are not enough training examples per unique foreign key value for $\textbf{R}_2$ in \textit{Yelp}.\footnote{Interestingly, 
the tuple ratio is similarly low ($2.6$) for $\textbf{R}_2$ in \textit{Books} but the error of \textit{NoJoin} is not much higher. So, the tuple 
ratio seems to be a \textit{conservative} indicator: it can tell if an error is likely to rise but the error may not actually rise in some cases.}
Almost every other dimension table can safely be discarded. A similar situation arises for the ANN on \textit{Yelp} and for the
 RBF-SVM on \textit{Yelp}, \textit{LastFM}, and \textit{Books}. 

Overall, out of $14$ dimension tables across the $7$ datasets that can potentially be discarded, we were able to safely discard $13$ for the 
decision tree and ANN, with the tuple ratio threshold being only about $3$x. For the RBF-SVM, we were able to discard $11$ dimension tables, 
with the tuple ratio threshold being about $6$x.
These results are surprising given the more conservative behavior predicted even for the linear classifiers in~\cite{hamlet}. For both Naive Bayes 
and logistic regression, only $7$ of the dimension tables were deemed ``safe to avoid" with the tuple ratio threshold being about $20$x. 
But of course, even tables that were predicted not safe to avoid could have been avoided without lowering accuracy significantly.
\textit{Overall, we see that the decision trees and ANN need six times \textit{fewer} training examples and the RBF-SVM needs three times fewer 
training examples than linear classifiers to avoid extra overfitting when avoiding the joins.} These results are counter-intuitive because conventional 
wisdom holds that such complex models need more (not fewer) training examples to avoid extra overfitting.

For an interesting comparison that we will use later on in Section 5, we also present the results for 1-NN (from package ``RWeka'' in R; it has no 
hyper-parameters). Surprisingly, as Table~\ref{Table:RealTest1} shows, the accuracy of even this braindead classifiers is sometimes comparable to 
decision trees and RBF-SVMs! More importantly, on most of the datasets, 1-NN with \textit{NoJoin} has a higher accuracy than with \textit{JoinAll}. 
We discuss this behavior further in Section 5.


\begin{figure*}[t]
\centering
\includegraphics[width=\linewidth]{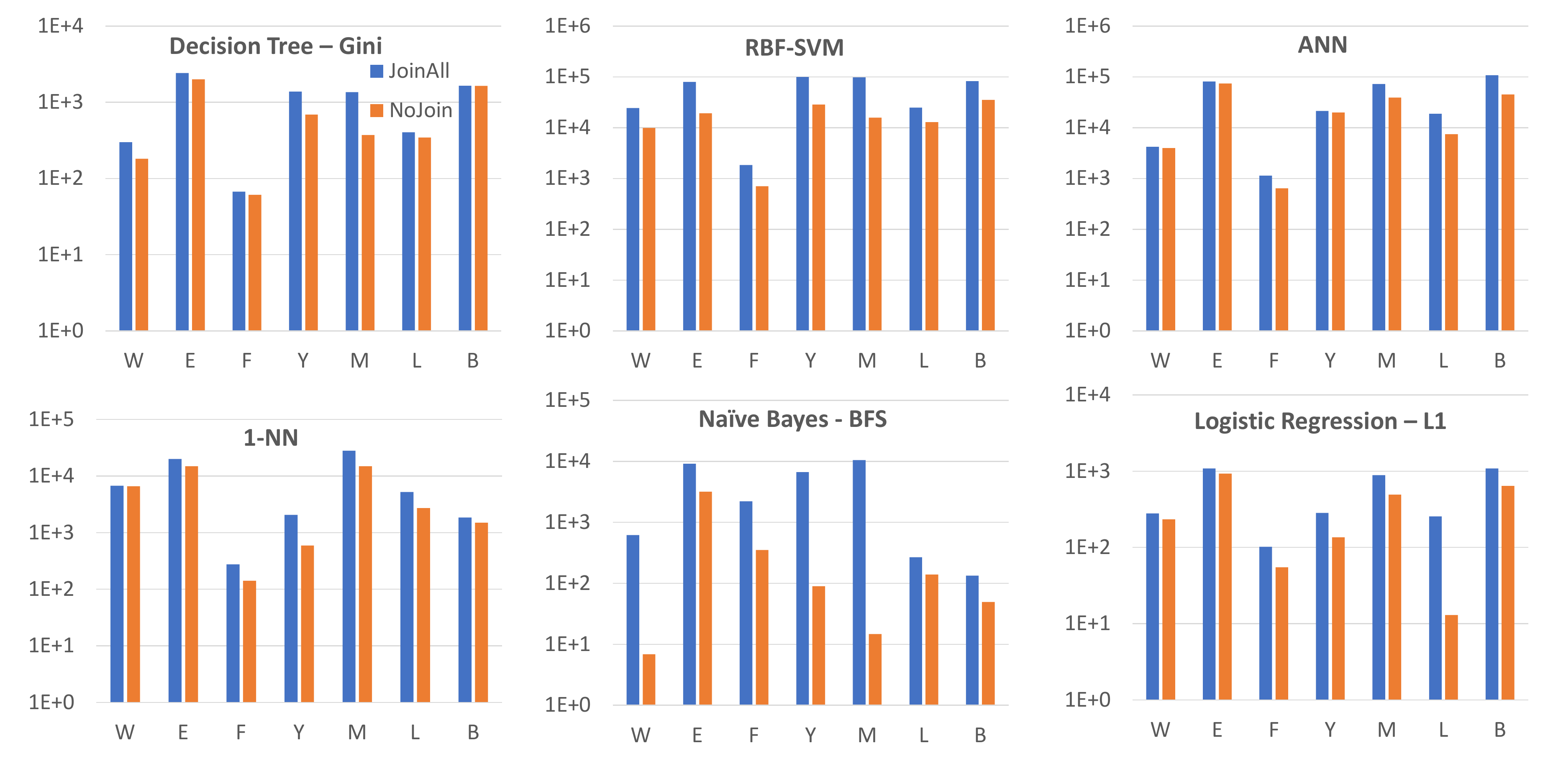}
\vspace{-5mm}
\caption{End-to-end runtimes on the real-world datasets: Walmart (W), Expedia (E), Flights (F), Yelp (Y), Movies (M), LastFM (L) and Books (B).}
\label{Figure:runtime}
\end{figure*}

\subsubsection*{Runtimes}
A key benefit of avoiding KFK joins safely is that ML runtimes (including feature selection) could be significantly lowered for the linear classifiers~\cite{hamlet}.
We now verify if this is the case for the high-capacity classifiers as well. We compare the runtimes of the end-to-end execution of the ML training (including the 
grid search) and testing for all models on all datasets.
Due to space constraints, we only report Gini metric for the decision tree and the RBF kernel for the SVM; these were also the most robust to avoiding joins.
All experiments (except for ANN) were run on CloudLab, which offers free access to physical compute nodes for research \cite{ricci2014introducing}. We use a custom OpenStack profile running Ubuntu 14.10 with 40 Intel Xeon cores and 160GB of RAM.
The ANN experiments were run on a commodity laptop with Nvidia GeForce GTX 1050 GPU, 16GB RAM and running Windows 10. The version of R used is 3.2.2 and the version of TensorFlow used is 1.1.0. Figure~\ref{Figure:runtime} presents the results.

For the high-capacity classifiers, we saw an average speedup of about 2x for \textit{NoJoin} over \textit{JoinAll}.  The highest speedup was 
on the \textit{Movies}: 3.6x for the decision tree and 6.2x for the RBF-SVM. As for the ANN, \textit{LastFM} reported the 
largest speedup of 2.5x. 
The speedup for the linear classifiers were more significant. For example, on \textit{Movies}, we saw a speedup of 707x for Naive Bayes, while on 
\textit{LastFM}, we saw a speedup of 20x for logistic regression. Thus, these results corroborate the orders of magnitude speedup reported in~\cite{hamlet} 
for Naive Bayes with backward selection.

\begin{figure*}[t]
\centering
\includegraphics[width=0.86\linewidth]{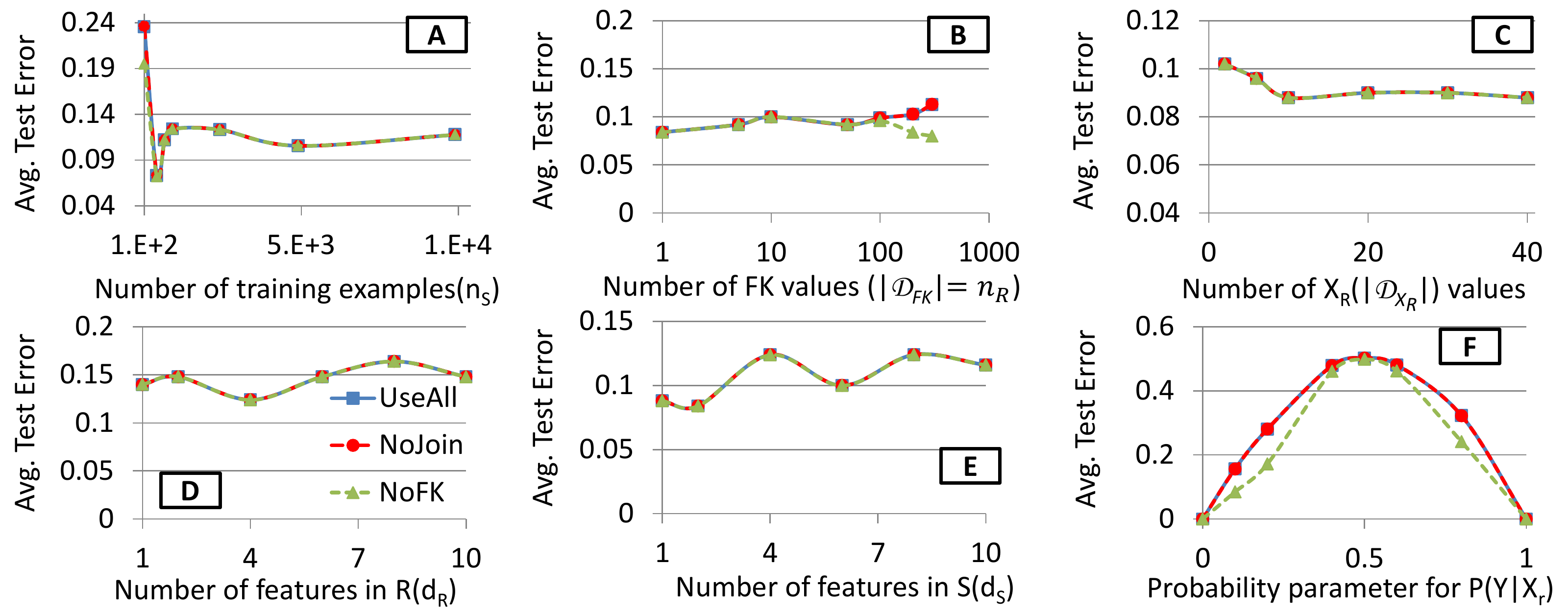}
\caption{Simulation results for Scenario \texttt{OneXr}. For all plots except (E), we fix $p = 0.1$. Note that $n_R \equiv |\mathcal{D}_{FK}|$.
(A) Vary $n_S$, while fixing $(n_R, d_S, d_R) = (40, 4, 4)$.
(B) Vary $n_R$, while fixing $(n_S, d_S, d_R) = (1000, 4, 4)$.
(C) Vary $d_S$, while fixing $(n_S, n_R, d_R) = (1000, 40, 4)$.
(D) Vary $d_R$, while fixing $(n_R, d_S, d_R) = (1000, 40, 4)$.
(E) Vary $p$,  while fixing $(n_S, n_R, d_S, d_R) = (1000, 40, 4, 4)$.
(F) Vary $|\mathcal{D}_{X_r}|$, while fixing $(n_S, n_R, d_S, d_R) = (1000, 40, 4, 4)$; all other features in $\textbf{X}_R$ and $\textbf{X}_S$ are binary.
}
\label{Figure:OneXrSimulation}
\end{figure*}

\section{In-depth Simulation Study}	

We now dive deeper into the behavior of the decision tree classifier using a simulation study in which we vary the properties of the 
underlying ``true'' data distribution. We focus on a two-table KFK join for simplicity. We sample datasets of different dimensions.
We use the decision tree for this study because it exhibited the maximum robustness to avoiding KFK joins on the real-world datasets.
Our simulation study is designed to comprehensively ``stress test'' this robustness.
Note that our simulation methodology is not tied to decision trees; it is generic enough to be applicable to classifier because we only 
use standard generic notions of error and \textit{net variance} as defined in~\cite{hamlet}.

\paragraph*{Setup and Data Synthesis}
There is one dimension table \textbf{R} ($q=1$), and all of $\textbf{X}_S$, $\textbf{X}_R$, and $Y$ are boolean (domain size $2$).
We control the ``true'' distribution $P(Y,\textbf{X})$ and sample labeled examples in an IID manner from it.
We study two different scenarios for what features are used to (probabilistically) determine $Y$: \texttt{OneXr} and \texttt{XSXR}.
These scenarios represent opposite extremes for how likely the (test) error is likely to shoot up when $\textbf{X}_R$ is discarded
and $FK$ is used as a representative~\cite{hamlet}. In \texttt{OneXr}, a lone feature $X_r \in \textbf{X}_R$ determines $Y$; the 
rest of $\textbf{X}_R$ and $\textbf{X}_S$ are random noise (but note that $FK$ will not 
be noise because it functionally determines $X_r$). In \texttt{XSXR}, all features in $\textbf{X}_S$ and $\textbf{X}_R$ determine $Y$. 
Intuitively, \texttt{OneXr} is the worst-case scenario for discarding $\textbf{X}_R$ because $X_r$ is typically 
far more succinct than $FK$, which we expect to translate to less possibility of overfitting with NoJoin. Note that if we use $FK$ 
directly in $P$, $\textbf{X}_R$ can be more easily discarded because $FK$ conveys more information anyway; so, we skip this scenario.

The following data parameters are varied one at a time: number of training examples ($n_S$), size of foreign key domain ($|\mathcal{D}_{FK}|$ $=n_R$),
number of features in $\textbf{X}_R$ ($d_R$), and number of features in $\textbf{X}_S$ ($d_S$). We also sample $\frac{n_S}{4}$ examples each for the 
validation set (for hyper-parameter tuning) and the holdout test set (final indicator of error).
We generate $100$ different training datasets and measure the average test error and average net variance (as defined in~\cite{pedrobvd}) 
based on the different models obtained from these $100$ runs.

\subsection{Scenario OneXr}

The ``true'' distribution is set as follows: $P(Y=0|X_r=0)=P(Y=1|X_r=1)=p$, where $p$ is called the probability skew parameter that controls the Bayes error (noise).
The exact procedure for sampling examples is as follows: (1) Construct tuples of \textbf{R} by sampling $\textbf{X}_R$ values randomly (each feature value 
is an independent coin toss). (2) Construct the tuples of \textbf{S} by sampling $\textbf{X}_S$ values randomly (independent coin tosses). (3) Assign $FK$ 
values to \textbf{S} tuples uniformly randomly from $\mathcal{D}_{FK}$. (4) Assign $Y$ values to \textbf{S} tuples by looking up into their respective $X_r$ 
value (implicit join on $FK = RID$) and sampling from the above conditional distribution.

We compare the same three approaches: \textit{JoinAll}, which uses $\textbf{X} \equiv [\textbf{X}_S, FK, \textbf{X}_R]$, 
\textit{NoJoin}, which uses $\textbf{X} \equiv [\textbf{X}_S, FK]$ (i.e., discard $\textbf{X}_R$), and \textit{NoFK}, 
which uses $\textbf{X} \equiv [\textbf{X}_S, \textbf{X}_R]$ (i.e., discard $FK$).
We include \textit{NoFK} for a lower bound on errors, since we know $FK$ does not determine determine $Y$ (although indirectly it 
does).\footnote{In general though, \textit{NoFK} could have much higher errors if $FK$ is part of the true distribution; indeed, \textit{NoFK} had 
much higher errors on many real datasets (Table~\ref{Table:RealTest1}).}
Figure~\ref{Figure:OneXrSimulation} presents the results for the (holdout) test errors for varying each relevant data and distribution parameter, one at a time.

Interestingly, regardless of the parameter being varied, in almost all cases, \textit{NoJoin} and \textit{JoinAll} have virtually identical errors (close to the Bayes 
error)! From inspecting the actual decision trees learned in these two settings, we found that in almost all cases, $FK$ was used heavily for partitioning and seldom was 
a feature from $\textbf{X}_R$, including $X_r$, used. This suggests that $FK$ can indeed act as a good representative of $\textbf{X}_R$ even in this extreme sccenario. 
In contrast to these results,~\cite{hamlet} reported that for linear models, the errors of \textit{NoJoin} shot up compared to \textit{JoinAll} (a gap of nearly $0.05$)
as the tuple ratio starts falling below $20$. In stark contrast, as Figure~\ref{Figure:OneXrSimulation}(B) shows, even for a tuple ratio of just $3$, \textit{NoJoin} 
and \textit{JoinAll} have similar errors with the decision tree. This corroborates the results seen for the decision tree on the real datasets (Table~\ref{Table:RealTest1}).
When $n_S$ becomes very low or when $|\mathcal{D}_{FK}|$ becomes very high, the absolute errors of \textit{JoinAll} and \textit{NoJoin} increase compared to \textit{NoFK}. 
This suggests that when the tuple ratio is very low, \textit{NoFK} is perhaps worth trying too. This is similar to the behavior seen on \textit{Yelp}. 
Overall, \textit{NoJoin} exhibits similar behavior as the current practice of \textit{JoinAll}.

\begin{figure}[t]
\centering
\includegraphics[width=0.99\linewidth]{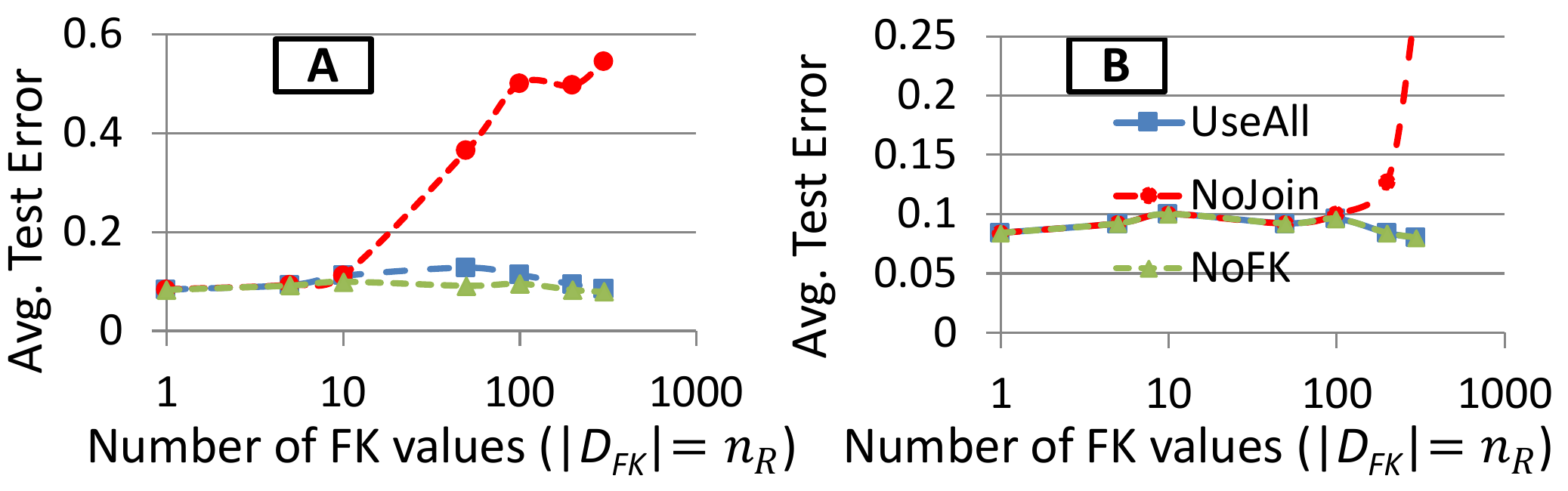}
\caption{Scenario \texttt{OneXr} simulations with the same setup as Figure~\ref{Figure:OneXrSimulation}(B), except for (A) 1-NN and (B) RBF-SVM.}
\label{Figure:OneXr1nnSVMSimulation}
\end{figure}

\begin{figure}[t]
\centering
\includegraphics[width=0.99\linewidth]{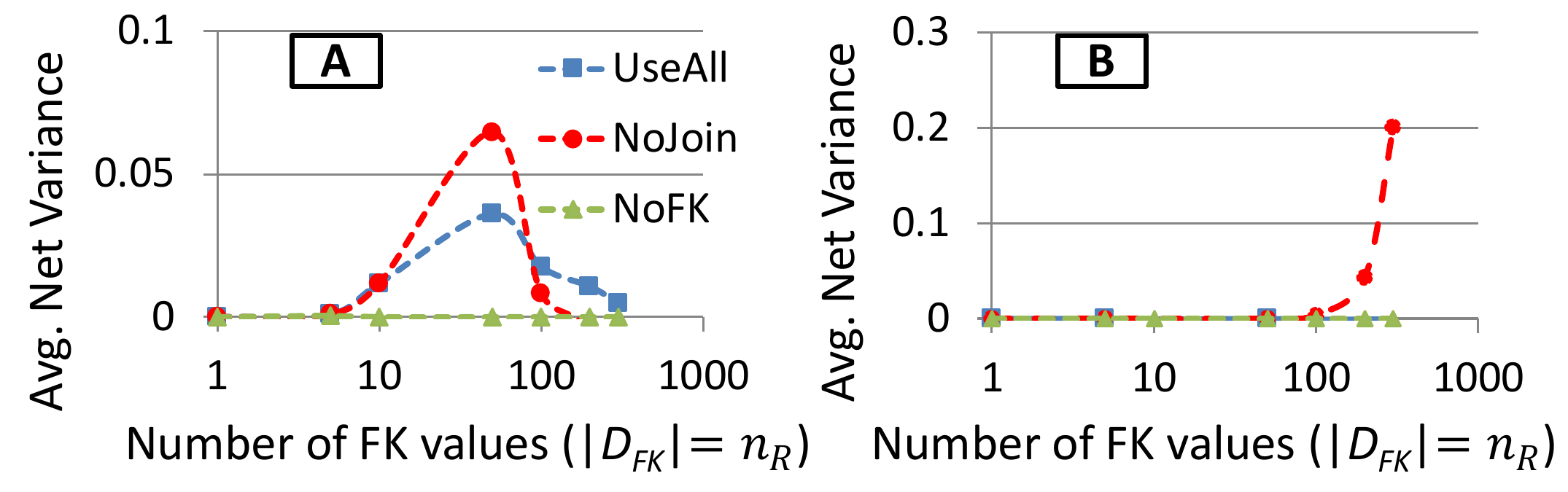}
\caption{Average net variance in the scenario \texttt{OneXr} for (A) 1-NN and (B) RBF-SVM.}
\label{Figure:OneXr1nnSVMSimulationVariance}
\end{figure}

\begin{figure*}[t]
\centering
\includegraphics[width=0.99\linewidth]{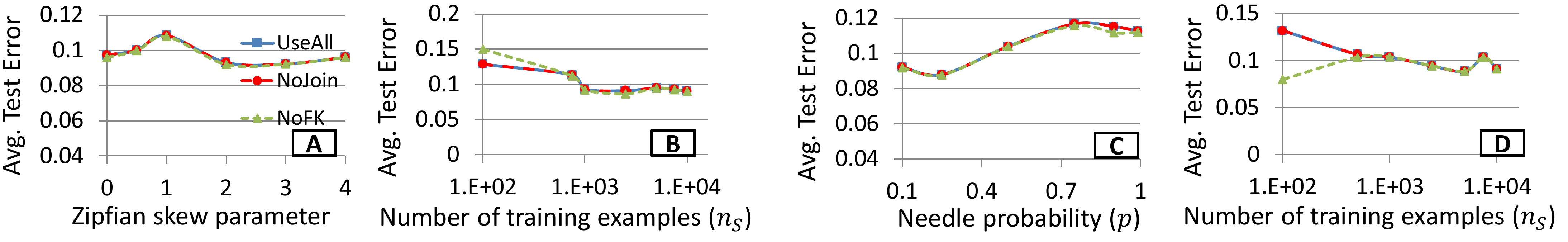}
\vspace{2mm}
\caption{Scenario \texttt{OneXr} simulations with skew in $P(FK)$. (A-B) Zipfian skew. (C-D) Needle-and-thread skew. For (A) and (C), we vary the respective skew parameter
(Zipfian skew parameter and needle probability), while fixing $(n_S, n_R, d_S, d_R) = (1000, 40, 4, 4)$. For (B) and (D), we vary $n_S$, 
while fixing $(n_R, d_S, d_R) = (40, 4, 4)$, the Zipfian skew parameter to $2$ for (B), and the needle probability to $0.5$ for (D).}
\label{Figure:OneXrZipfSimulation}
\end{figure*}

Finally, we also ran this scenario for the RBF-SVM (and 1-NN) and found the trends to be similar, except for the magnitude of the tuple ratio at which 
\textit{NoJoin} deviates from \textit{JoinAll}. Figure~\ref{Figure:OneXr1nnSVMSimulation} presents the results for the experiment in which we increase $|\mathcal{D}_{FK}|=n_R$, 
while fixing everything else, similar to Figure~\ref{Figure:OneXrSimulation}(B) for the decision tree. We see that for the RBF-SVM, the error deviation starts when the 
tuple ratio ($n_S/n_R$) falls below roughly $6$x. This corroborates its behavior on the real datasets (Table~\ref{Table:RealTest2}).
The 1-NN, as expected, is far less stable and the deviation starts even at a tuple ratio of $100$x, i.e., $n_R = 10$).
As Figure~\ref{Figure:OneXr1nnSVMSimulationVariance} confirms, the deviation in accuracy for the RBF-SVM arises due to the net variance, 
which helps quantify the extra overfitting. This is akin to the extra overfitting reported in~\cite{hamlet} using the plots of the net variance.
Intriguingly, the 1-NN sees its net variance exhibit non-monotonic behavior; this is likely an artifact of its unstable behavior, since fewer and fewer
training examples will match on $FK$ as $n_R$ keeps rising.

\begin{figure*}[t]
\centering
\includegraphics[width=0.99\linewidth]{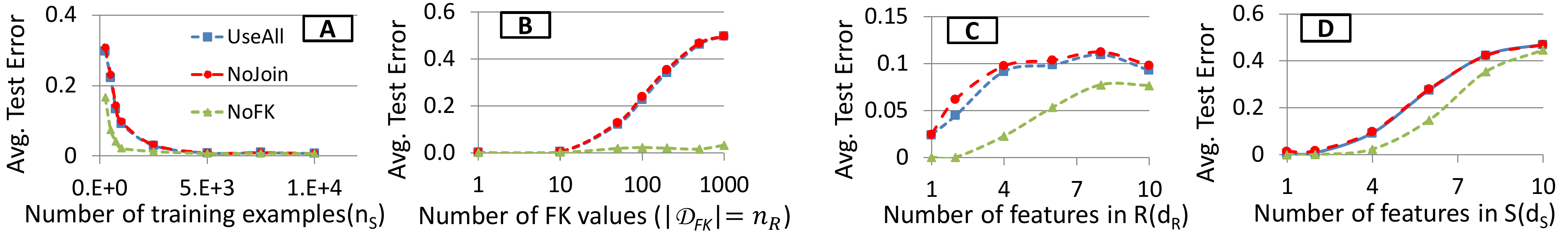}
\vspace{2mm}
\caption{Simulation results for Scenario \texttt{XSXR}. The parameter values varied/fixed ($n_S$, $n_R$, $d_S$, and $d_R$) are the same as in Figure~\ref{Figure:OneXrSimulation} (A)-(D).}
\label{Figure:XsXrSimulation}
\end{figure*}

\paragraph*{Foreign Key Skew}
The regular \texttt{OneXr} scenario samples $FK$ uniformly randomly from $\mathcal{D}_{FK}$ (step 3 in the procedure). We now ask if a \textit{skew} in 
the distribution of $FK$ values could widen the gap between \textit{JoinAll} and \textit{NoJoin}. To study this scenario, we modify the data generation procedure slightly:
in step 3, we sample $FK$ values with a Zipfian skew or a needle-and-thread skew. The Zipfian skew simply uses a Zipfian distribution for $P(FK)$ controlled by the Zipfian
skew parameter. The needle-and-thread skew allocates a large probability mass (parameter $p$) to a single $FK$ value (the ``needle'') and uniformly distributes the rest of 
the probability mass to all other $FK$ values (the ``thread''). For the linear model case,~\cite{hamlet} reported that as the skew parameters increased, the gap widened.
Figure~\ref{Figure:OneXrZipfSimulation} presents the results for the decision tree.

Surprisingly, the gap between \textit{NoJoin} and \textit{JoinAll} does not widen significantly no matter how much skew introduced in either the Zipfian or the 
needle-and-thread case! This result further affirms the remarkable robustness of the decision tree when discarding foreign features. As expected, 
\textit{NoFK} is better when $n_S$ is low, while overall, \textit{NoJoin} is quite similar to \textit{JoinAll}.

\subsection{Scenario XSXR}

Unlike the \textit{OneXr} scenario, we now create a true distribution that maps $\textbf{X} \equiv [\textbf{X}_S, \textbf{X}_R]$ to $Y$ without any Bayes error (noise).
The exact procedure for sampling examples is as follows: (1) Construct a true probability table (TPT) with entries for all possible values of 
$[\textbf{X}_S, \textbf{X}_R]$ and assign a random probability to each entry such that the total probability is $1$.
(2) For each entry in the TPT, pick a $Y$ value randomly and append the TPT entry; this ensures $H(Y|\textbf{X}) = 0$.
(3) Marginalize the TPT to obtain $P(\textbf{X}_R)$ and from it, sample $n_R = \mathcal{D}_{FK}$ tuples for \textbf{R} along with an associated sequential $RID$ value.
(4) In the original TPT, push the probability of each entry to $0$ if its $\textbf{X}_R$ values did not get picked for \textbf{R} in step 3.
(5) Renormalize the TPT so that the total probability is $1$ and sample $n_S$ examples ($Y$ values do not change) and construct \textbf{S}.
(6) For each tuple in \textbf{S}, pick its $FK$ value uniformly randomly from the subset of $RID$ values that map to its $\textbf{X}_R$ value in \textbf{R} (an implicit join).

We compare three settings: \textit{JoinAll}, \textit{NoJoin}, and \textit{NoFK}, with \textit{NoFK} meant to be a lower bound on the errors possible (because 
it uses the knowledge that $FK$ is not directly a part of the true distribution). Once again, our hypothesis is that \textit{JoinAll} and \textit{NoJoin} will exhibit similar 
errors in most cases, while \textit{NoFK} will perform better when the tuple ratio is low. Figure~\ref{Figure:XsXrSimulation} presents the results.

Once again, we see that \textit{NoJoin} and \textit{JoinAll} exhibit similar errors in almost all cases, with the largest gap being $0.017$ in Figure~\ref{Figure:XsXrSimulation}(C)).
Interestingly, even when the tuple ratio is close to $1$, the gap between \textit{NoJoin} and \textit{JoinAll} does not widen much. 
Figure~\ref{Figure:XsXrSimulation}(B)) shows that as $|\mathcal{D}_{FK}|$ increases, \textit{NoFK} remains at low overall errors, unlike both \textit{JoinAll} and \textit{NoJoin}.
But as we increase $d_R$ or $d_S$, the gap between \textit{JoinAll}/\textit{NoJoin} and \textit{NoFK} narrows because even \textit{NoFK} does not have enough training examples.
Of course, all gaps virtually disappear as the number of training examples increases, as shown by Figure~\ref{Figure:XsXrSimulation}(A).
Overall, \textit{NoJoin} exhibits similar behavior as the current practice of \textit{JoinAll} even in this scenario.

\begin{figure}[t]
\centering
\includegraphics[width=0.99\linewidth]{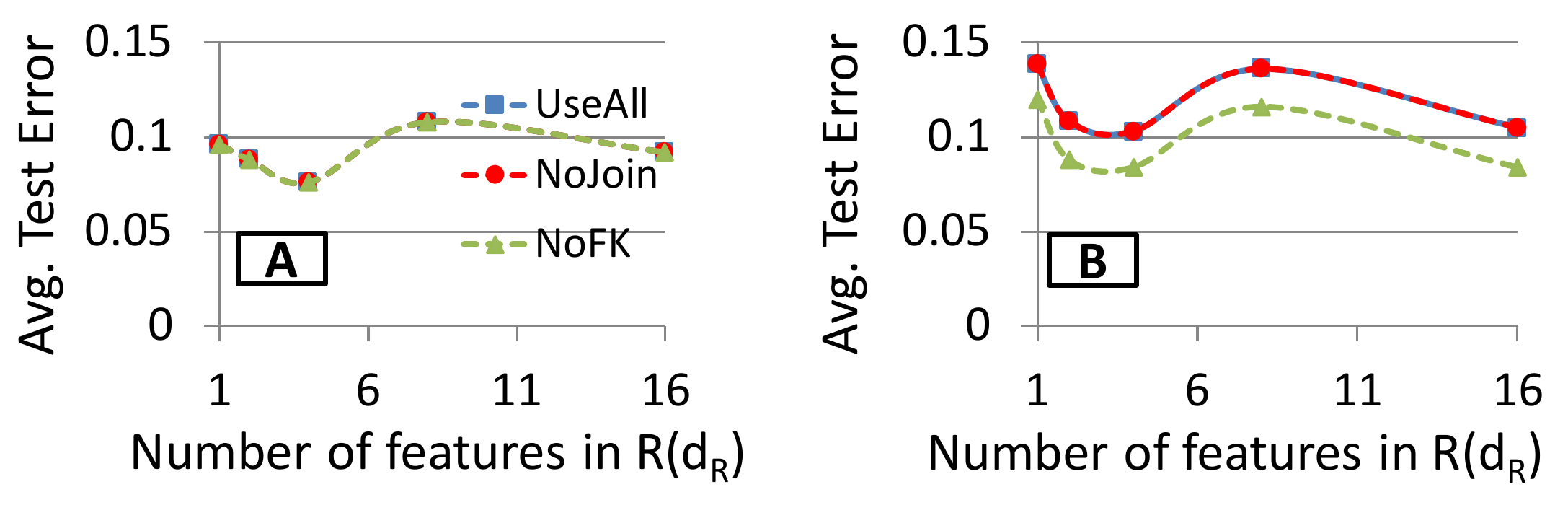}
\caption{Scenario \texttt{RepOneXr} simulations for decision tree. (A) Vary $d_R$ while fixing $(n_S, n_R, d_S) = (1000, 40, 4)$. 
(B) Vary $d_R$ while fixing $(n_S, n_R, d_S) = (1000, 200, 4)$.}
\label{Figure:OneXrjerry_dt}
\end{figure}

\begin{figure}[t]
\centering
\includegraphics[width=0.99\linewidth]{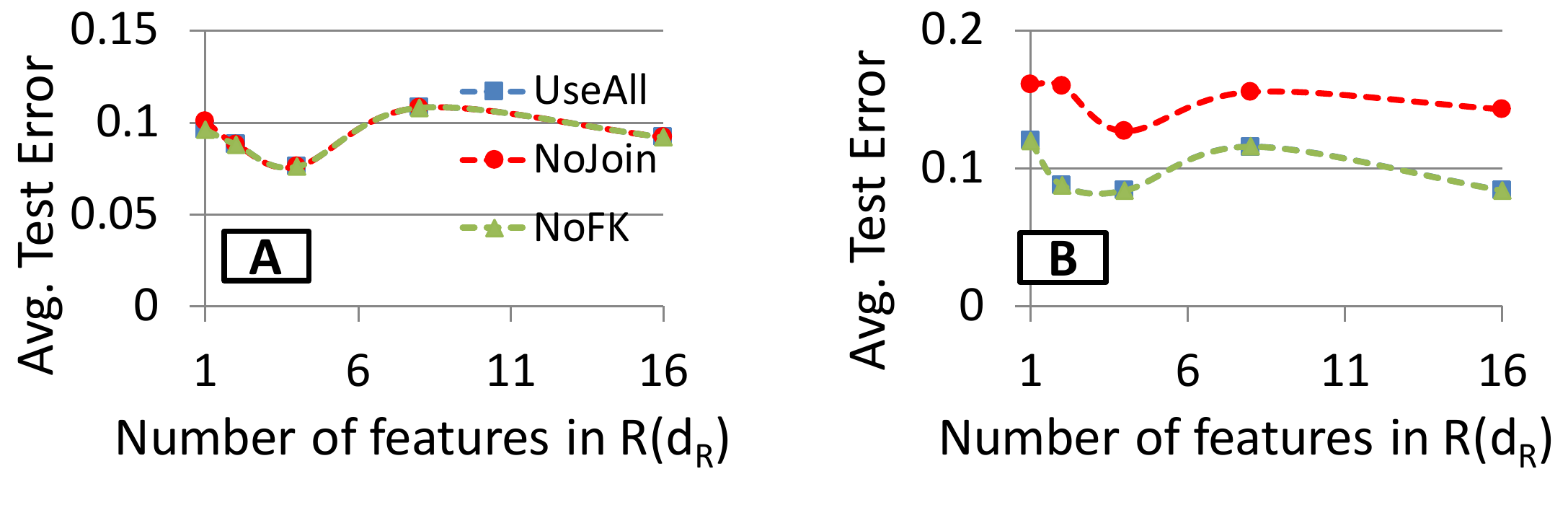}
\caption{Scenario \texttt{RepOneXr} simulations with same setup as Figure~\ref{Figure:OneXrjerry_dt}, except for RBF-SVM.}
\label{Figure:OneXrjerry_svm}
\end{figure}

\begin{figure}[t]
\centering
\includegraphics[width=0.99\linewidth]{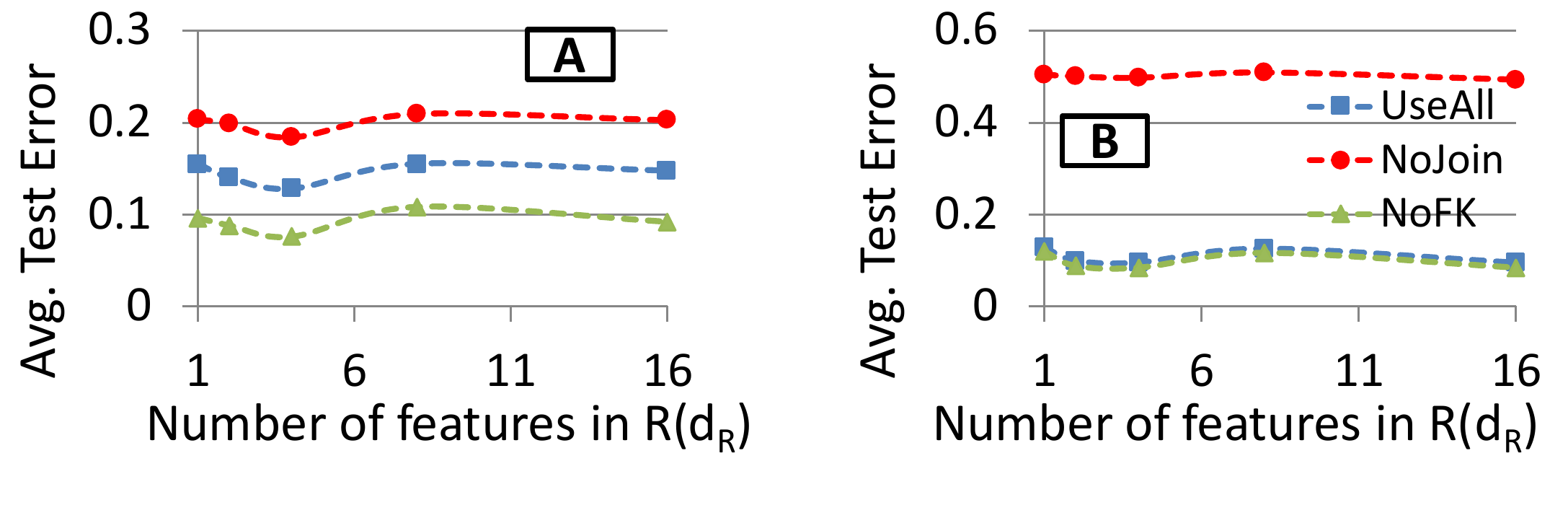}
\caption{Scenario \texttt{RepOneXr} simulations with same setup as Figure~\ref{Figure:OneXrjerry_dt}, except for 1-NN.}
\label{Figure:OneXrjerry_1nn}
\end{figure}

\subsection{Scenario RepOneXr}
We now present results for a new simulation scenario in which the true distribution is precisely captured using a lone feature $X_r \in \textbf{X}_R$. We sample examples similarly as per the procedure mentioned earlier for \texttt{OneXr}, except that the tuples of $\textbf{R}$ will now be constructed by replicating the value of $X_r$ sampled for a tuple to create all the other features in $X_R$. That is, $X_R$ of an example is just the same value repeated $d_R$ times. Note that the FD $\textit{FK} \rightarrow X_R $ implies there are at least as many unique $FK$ values as $\textbf{X}_R$ values. Thus, by increasing the number of $FK$ values relative to $X_R$ values, we hope to increase the chance of the model getting ``confused'' with \textit{NoJoin}. Our goal is to see if this widens the gap between \textit{JoinAll} and \textit{NoJoin}.

Figure~\ref{Figure:OneXrjerry_dt} presents the results for the two experiments on decision trees where (A) has a high tuple ratio of 25x and (B) has a low tuple ratio of 5x. 
We see that once again, \textit{JoinAll} and \textit{NoJoin} exhibit similar errors in both the cases. We also run this simulation scenario for 
both the RBF-SVM and 1-NN as well; the results are shown in Figure~\ref{Figure:OneXrjerry_svm}) and Figure~\ref{Figure:OneXrjerry_1nn} respectively.
We see that the trends are similar to the decision tree. For the RBF-SVM, the error of \textit{NoJoin} deviates at a tuple ratio of about 5x. 
As for 1-NN, as expected, it is much less stable and the deviation happens even at a higher tuple ratio of 25x.
At low tuple ratios, as expected, the absolute errors of \textit{JoinAll} and \textit{NoJoin} increase compared to \textit{NoFK} for all three models.

\section{Analysis and Open Questions}

\subsection{Explaining the Results}

We now intuitively explain the surprising behavior of decision trees and RBF-SVM with \textit{NoJoin} vis-a-vis \textit{JoinAll}.
We first ask: {Does \textit{NoJoin} compromise the ``generalization error''?} The generalization error is the difference of the test and train errors.
Tables~\ref{Table:RealTest1} and~\ref{Table:RealTest2} already provided the test accuracy. 
Tables~\ref{Table:RealTrain1} and~\ref{Table:RealTrain2} now provide the train accuracy. Clearly, \textit{JoinAll} vs 
\textit{NoJoin} are almost indistinguishable for the decision tree! The only exception is \textit{Yelp}, which we already noted. 
Note that the absolute generalization error is often high, which is expected for decision trees~\cite{dtreebias2}. 
For example, the train accuracy is nearly $100\%$ on \textit{Flights}, while the test accuracy on it is only $85\%$.
But the absolute generalization error is orthogonal to our focus; we only note that \textit{NoJoin} does not increase this difference significantly.
\textit{In other words, discarding foreign features did not significantly affect the generalization error of the decision tree!} 
The generalization errors of the RBF-SVM also exhibit a similar trend.

Returning to 1-NN, Table~\ref{Table:RealTest1} showed that it has similar accuracy as RBF-SVM on some datasets. 
We now explain why that comparison is useful: the RBF-SVM essentially behaves similar to the 1-NN in some cases 
when $FK$ is used (both \textit{JoinAll} and \textit{NoJoin})! But this does not necessarily hurt its test accuracy.
Note that $FK$ is represented using the standard one-hot encoding for RBF-SVM and 1-NN. So, $FK$ can contribute 
to a maximum distance of $2$ in a (squared) Euclidean distance between two examples $x_i$ and $x_j$.
But since $\textbf{X}_R$ is functionally dependent on $FK$, if $x_i.FK = x_j.FK$, then $x_i.\textbf{X}_R = x_j.\textbf{X}_R$. So, if $x_i.FK = x_j.FK$, the only contributor to 
the distance is $\textbf{X}_S$. But in many of the datasets, since $\textbf{X}_S$ is empty ($d_S = 0$), $FK$ becomes the sole determiner of the distances for \textit{NoJoin}.
This is akin to sheer \textit{memorization} of a feature's large domain. Since we operate on features with finite domains, test examples will also have $FK$ from that domain. 
Thus, memorizing $FK$ does not hurt generalization. While this seems similar to how deep neural networks excel at sheer memorization but still offer good test accuracy~\cite{rechtdnn}, 
the models in our setting are not necessarily memorizing all features -- only the foreign keys.
A similar explanation holds for the decision tree. If $\textbf{X}_S$ is not empty, then it will likely play a major role in the distance computations and our setting 
becomes more similar to the traditional single-table learning setting (no FDs).

\begin{table*}[t]
\centering
\includegraphics[width=0.99\linewidth]{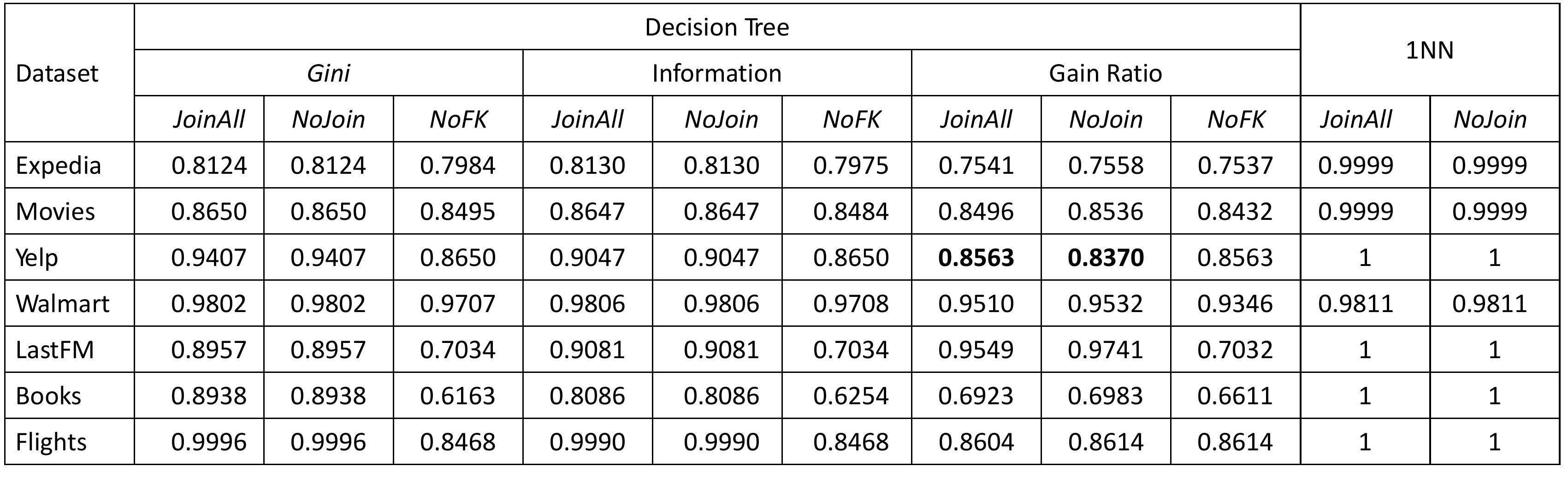}
\caption{Training accuracy for the same experiments as in Table~\ref{Table:RealTest1}.
Our goal is \textit{not} to compare the accuracy across ML models, but rather compare
the accuracy of \textit{JoinAll} and \textit{NoJoin} within each model.
The bold font marks the only cases where the accuracy of \textit{NoJoin} is at least 1\% lower 
than the accuracy of \textit{JoinAll}.
}
\label{Table:RealTrain1}
\end{table*}

\begin{table*}[t]
\centering
\includegraphics[width=0.99\linewidth]{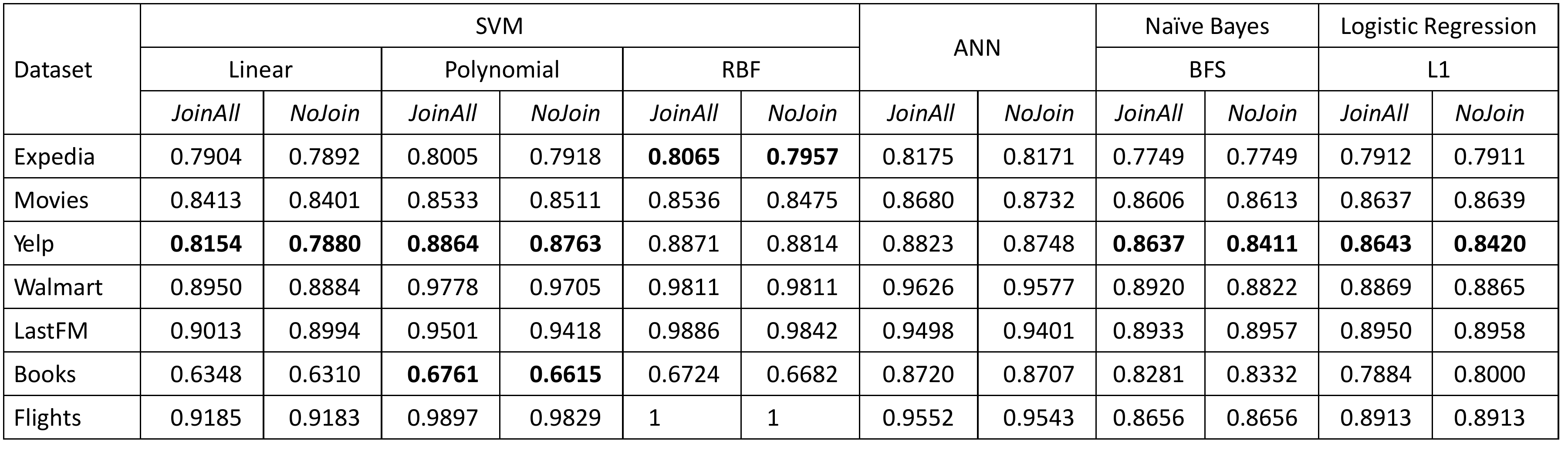}
\caption{Training accuracy for the same experiments as in Table~\ref{Table:RealTest2}.
Our goal is \textit{not} to compare the accuracy across ML models, but rather compare
the accuracy of \textit{JoinAll} and \textit{NoJoin} within each model.
The bold font marks the only cases where the accuracy of \textit{NoJoin} is at least 1\% lower 
than the accuracy of \textit{JoinAll}.
}
\label{Table:RealTrain2}
\end{table*}

We now explain why \textit{NoJoin} might deviate from \textit{JoinAll} when the tuple ratio is very low for the RBF-SVM. Even if $x_i.FK \ne x_j.FK$, it is possible that 
$x_i.\textbf{X}_R = x_j.\textbf{X}_R$. Suppose the ``true'' distribution is captured by $\textbf{X}_R$, e.g., as in OneXr. 
If the tuple ratio is very low, there are many $FK$ values but the number of $\textbf{X}_R$ values might still be small. In this case, given $x_i$, RBF-SVM (and 1-NN) is more 
likely to pick an $x_j$ that minimizes the distances on $\textbf{X}_R$, thus potentially yielding lower errors. But since \textit{NoJoin} does not have access to 
$\textbf{X}_R$, it can only use $\textbf{X}_S$ and $FK$. So, if $\textbf{X}_S$ is mostly noise, the possibility of the model getting ``confused'' 
increases. To see why, if there are very few other examples that share $x_i.FK$, then matching on $\textbf{X}_S$ might become more important. 
Thus, a non-match on $FK$ becomes more likely, which means a non-match on the implicit $\textbf{X}_R$ becomes more likely, which in turns makes higher errors more likely. 
But if there are more examples that share $x_i.FK$, then a match on $FK$ is more likely. Thus, as the tuple ratio increases, the gap between \textit{NoJoin} 
and \textit{JoinAll} disappears, as Figure~\ref{Figure:OneXr1nnSVMSimulation} showed. Internally, the RBF-SVM seems more robust to such chance mismatches by learning a higher-level 
relationship between all features compared to the stark 1-NN. Thus, the RBF-SVM is more robust to discarding foreign features at lower tuples ratios than 1-NN.

Finally, focusing on the decision tree, its internal feature selection and partitioning seems to make it quite robust to noise from any other features. Suppose again the ``true'' 
distribution is similar to OneXr. Since $FK$ already encodes all information that $\textbf{X}_R$ provides~\cite{hamlet}, the tree almost always uses $FK$ in its partitioning, 
often multiple times. This is not necessarily ``bad'' for test accuracy because test examples share the $FK$ domain. 
But when the tuple ratio becomes extremely low, the chance of $\textbf{X}_S$ ``confusing'' the tree over the information $FK$ provides goes up, potentially 
leading to higher errors with \textit{NoJoin}. \textit{JoinAll} escapes such a confusion thanks to $\textbf{X}_R$. If $\textbf{X}_S$ is empty, then $FK$ will almost surely
be used for partitioning. But with very few training examples per $FK$ value, the chance of sending it to a wrong partition goes up, leading to higher errors. It turns out 
that even with just $3$ or $4$ training examples per $FK$ value, such issues get mitigated. Thus, the decision tree seems even more robust to discarding foreign features.

\eat{
\begin{figure}[t]
\centering
\includegraphics[width=\columnwidth,height=\textheight,keepaspectratio]{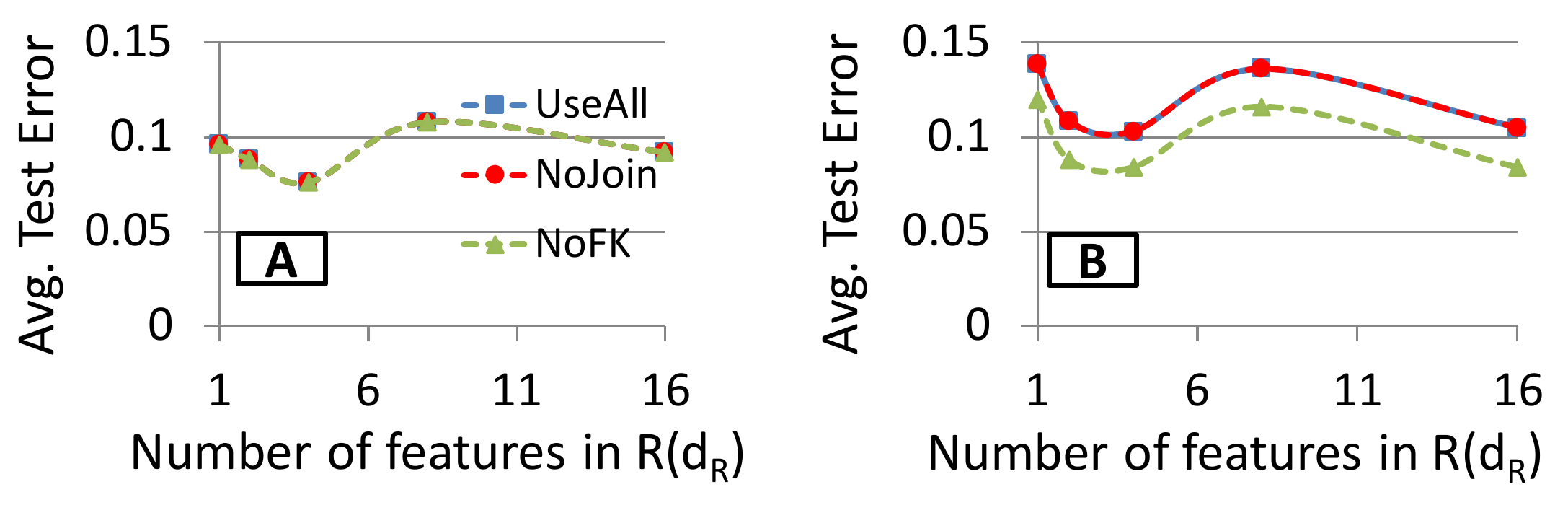}
\includegraphics[width=\columnwidth,height=\textheight,keepaspectratio]{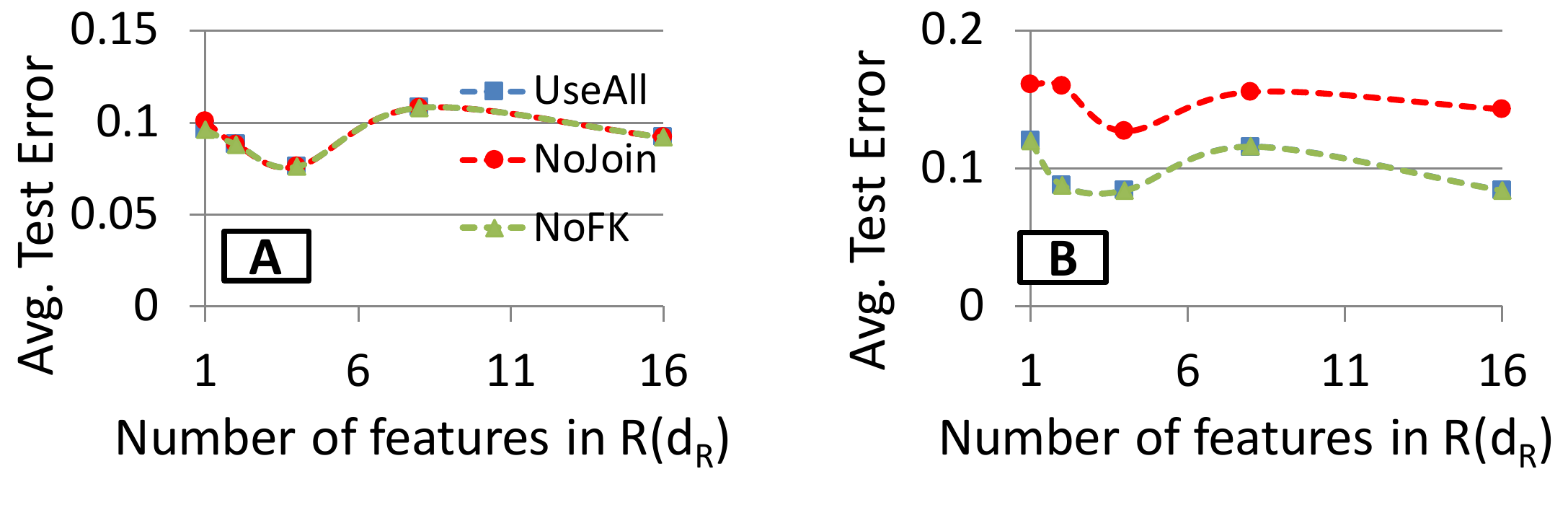}
\includegraphics[width=\columnwidth,height=\textheight,keepaspectratio]{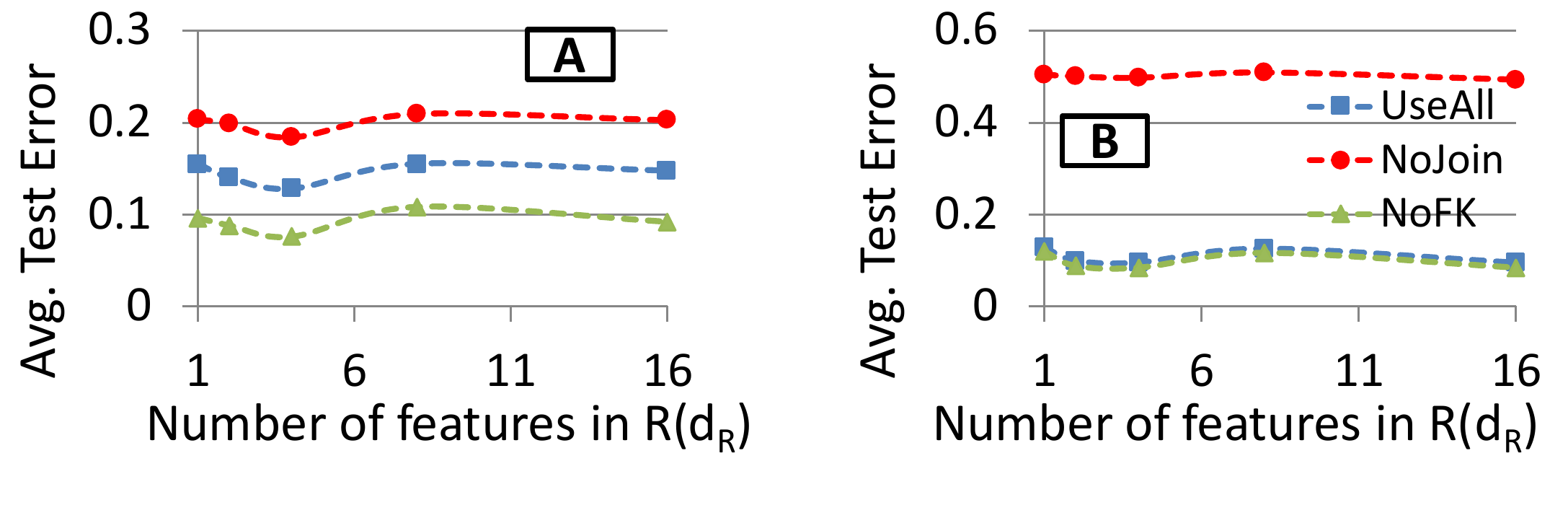}
\caption{OneXr simulation for repeated Xr Features}
\label{Figure:OneXrjerry}
\end{figure}

\paragraph*{Scenario RepeatedOneXr}
}

\subsection{Open Research Questions}

While our intuitive explanations capture the fine-grained behavior of the decision tree and RBF-SVM with \textit{NoJoin} vis-a-vis \textit{JoinAll}, there are many 
open questions for more research. Is it possible to quantify the probability of wrong partitioning with a decision tree as a function of the data properties?
Is it possible to quantify the probability of mismatched examples being picked by the RBF-SVM? Why does the theory of VC-dimensions predict 
the opposite of the observed behavior with these models? How do we quantify their generalizability if memorization is allowable and what forms of memorization are allowed?
Answering these questions would provide deeper insights into the effects of KFKDs/FDs on the generalizability and accuracy of such classifiers. It could also yield more 
formal mechanisms to characterize when discarding foreign features is feasible beyond just looking at tuple ratios.

From a data management perspective, there are database dependencies more general than FDs: 
embedded multi-valued dependencies (EMVDs) and join dependencies (JDs)~\cite{avibook}. How does the presence of such 
database dependencies among features affect the behavior of ML models? There are also conditional FDs, which satisfy FD-like 
constraints among subsets of the dataset~\cite{avibook}. How do such properties of the data distribution affect ML behavior?
Finally, the axioms of FDs imply that foreign features can be divided into arbitrary subsets before being avoided, which opens 
up a new trade-off space between fully avoiding a foreign table and fully using it. How do we quantify this trade-off and exploit it?
Answering these questions could open up new connections between data management and ML theory and potentially lead to 
new functionalities for ML systems.

\section{Making FK Features Practical}

We now discuss two key practical issues caused by the large domains of foreign key features and verify how standard approaches can be adapted to resolve them. 
In contrast to prior work on handling regular large-domain features~\cite{dtreebias1}, foreign key features are distinct in that they have coarser-grained side 
information available in the form of foreign features. This suggests an alternative way to exploiting such features, if possible, rather than always joining them in.

\subsection{Foreign Key Domain Compression}

While foreign keys often act as good representatives of foreign features for \textit{accuracy},
they pose a practical bottleneck for \textit{interpretability} due to their domain sizes.
For example, it is hard to visualize a decision tree that uses a foreign key feature with 1000s of values.
In order to make foreign key features more practical, we consider a simple approach that is standard in the ML literature:
\textit{lossy domain compression} to a (much) smaller user-defined domain size. Essentially, given a 
foreign key feature $FK$ with domain $\mathcal{D}_{FK}$ recoded as $[m]$ (where $m = |\mathcal{D}_{FK}|$) and a user-specified 
positive integer ``budget'' $l \ll m$, we want a mapping $f: [m] \rightarrow [l]$.

\begin{figure}[t]
\centering
\includegraphics[width=0.99\linewidth]{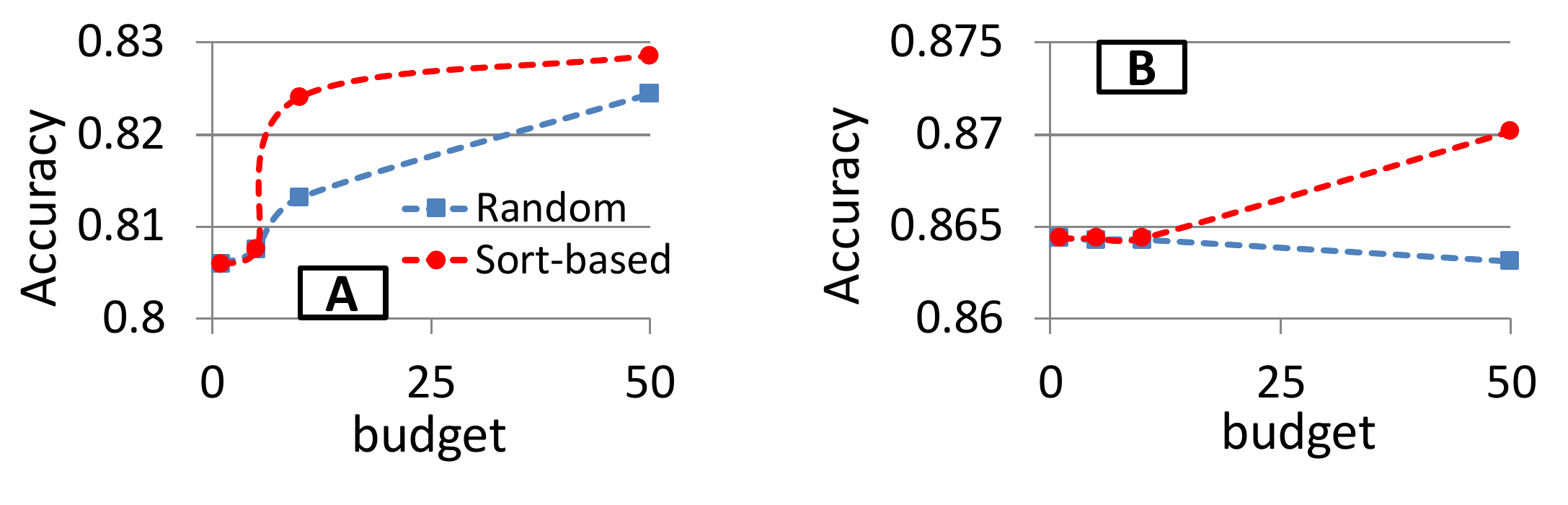}
\caption{Domain compression. (A) \textit{Flights}. (B) \textit{Yelp}.}
\label{Figure:Compression}
\end{figure}

A standard unsupervised method to construct $f$ is the \textit{Random hashing} trick~\cite{hashingtrick}, i.e., randomly hash from $[m]$ to $[l]$.
We also try a simple supervised method we call the \textit{Sort-based} method to preserve more of the information contained 
in $FK$ about $Y$. Sort-based is a greedy approach: sort $\mathcal{D}_{FK}$ based on $H(Y|FK=z), ~z \in \mathcal{D}_{FK}$, compute the 
differences among adjacent pairs of values, and pick the boundaries corresponding to the top $l-1$ differences (ties broken randomly). 
This gives us an $l$-partition of $\mathcal{D}_{FK}$. The intuition is that by grouping $FK$ values that have comparable conditional entropy, 
$H(Y|f(FK))$ is unlikely to be much higher than $H(Y|FK)$. Note that the lower $H(Y|FK)$ is, the more informative $FK$ is to predict $Y$. 
We leave more sophisticated approaches to future work.

We now empirically compare the above two heuristics using two of the real datasets for the Gini decision tree with \textit{NoJoin}. Our methodology 
is as follows. We retain the 50:25:25 train-validate-test split from before. We use the training split to construct $f$ and then compress $FK$ for 
the whole dataset. We then use the validation set as before to tune the hyper-parameters and measure the holdout test error. For random hashing, we 
report the average across five runs. Figure~\ref{Figure:Compression} presents the results.

On \textit{Yelp}, both \textit{Random} and \textit{Sort-based} have comparable accuracy although Sort-based is marginally higher, especially
as the budget $l$ increases. But on \textit{Flights}, we see a larger gap for some values of $l$ although the gap narrows as the $l$ increases.
The test accuracy with the whole $\mathcal{D}_{FK}$ ($l=m$) for \textit{NoJoin} on \textit{Flights} was $0.8516$ (see Table~\ref{Table:RealTest1}). So, 
it is surprising the test accuracy is about $0.83$ with such high compression. Even more surprisingly, the test accuracy with the whole 
$\mathcal{D}_{FK}$ ($l=m$) for \textit{NoJoin} on \textit{Yelp} was $0.8204$ and for \textit{NoFK} was $0.8644$. So, with domain compression,
we see significantly higher accuracy for \textit{NoJoin}, even higher than \textit{NoFK}. Overall, these results suggest that $FK$ domain compression 
is a promising way to resolve the large-domain issue rather than simply dropping $FK$.

\subsection{Foreign Key Smoothing}

Another issue caused by large $|\mathcal{D}_{FK}|$ is that some $FK$ values might not arise in the train set but arise 
in the test set or during deployment. This is not a cold start issue -- the $FK$ values are all still from the fully known $\mathcal{D}_{FK}$. 
This issue arises because there are not enough labeled examples to cover all of $\mathcal{D}_{FK}$ during training. 
Typically, this issue is handled using some form of \textit{smoothing}, e.g., Laplacian smoothing
for Naive Bayes by adding a pseudocount of $1$ to all frequency counts~\cite{mitchellbook}.
While similar smoothing techniques have been studied for probability estimation using decision trees~\cite{pedro2003}, to the best of our knowledge, 
this issue has not been handled in general for classification using decision trees. In fact, popular decision tree implementations in R simply 
crash if a value of $FK$ not seen during training arises during testing! Note that SVMs (or any other classifier operating on numeric 
feature spaces) do not face this issue due to the one-hot encoding of $FK$. 

We consider a simple approach to mitigate this issue: smooth by \textit{reassigning} an $FK$ value not seen during training to an $FK$ value that was
seen. There are various ways to reassign; for simplicity sake, we only study two lightweight unsupervised methods.
We leave more sophisticated approaches to future work.
We consider both \textit{random} reassignment and alternative approach that uses the foreign features ($\textbf{X}_R$) to decide the reassignment. 
Note that the latter is only feasible in cases where the dimension tables are available and not discarded. Since \textbf{R} provides auxiliary 
descriptive information about $FK$, we can utilize it for smoothing even if not for learning directly over them.
Our algorithm is simple: given a test example with $FK$ not seen during training, obtain an $FK$ seen during training whose corresponding 
$\textbf{X}_R$ feature vector has the minimum $l_0$ distance with the test example's $\textbf{X}_R$ (ties broken randomly). The $l_0$ distance is
simply the count of the number of pairwise mismatches of the respective features in the two $\textbf{X}_R$ feature vectors. 

\begin{figure}[t]
\centering
\includegraphics[width=0.99\linewidth]{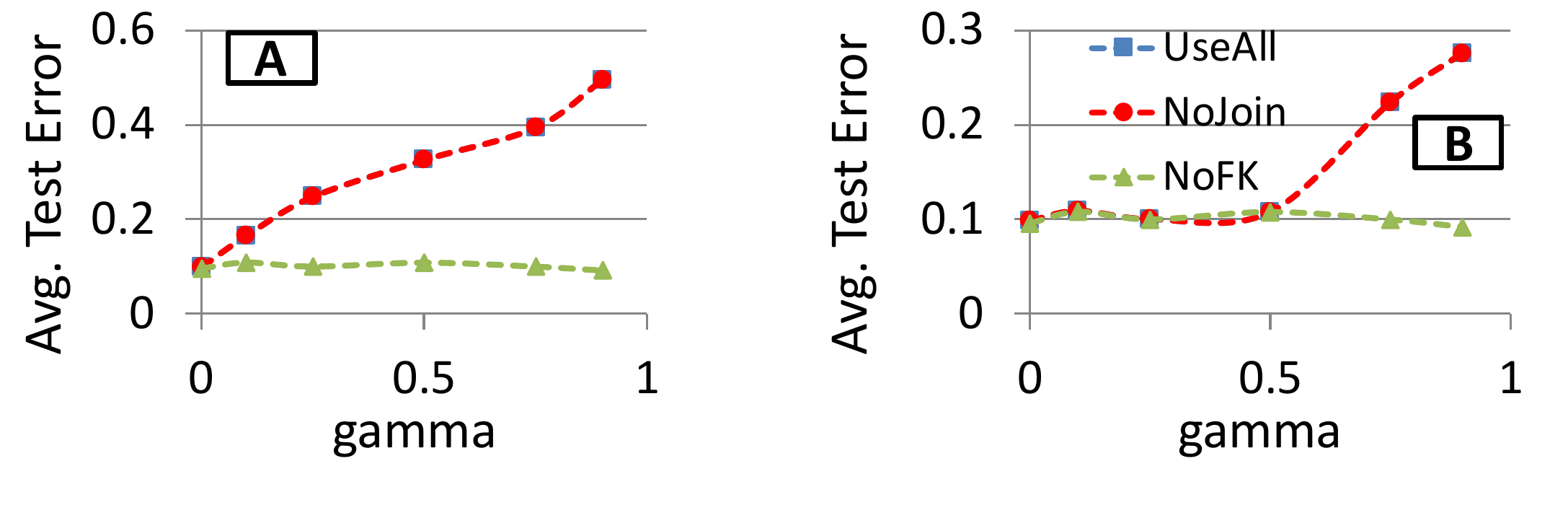}
\caption{Smoothing. (A) Hashing. (B) $\textbf{X}_R$-based.}
\label{Figure:smoothing}
\end{figure}

The intuition for $\textbf{X}_R$-based smoothing is that if $\textbf{X}_R$ is part of the ``true'' distribution, it may yield higher accuracy 
than random reassignment. But if $\textbf{X}_R$ is just noise, this becomes essentially random reassignment.
To validate our claim, we use the OneXr simulation scenario. Recall that a feature $X_r \in \textbf{X}_R$
determines the target (with some Bayes noise as before). We introduce a parameter $\gamma$ that is the ratio of the number of $FK$ values not seen 
during training to $|\mathcal{D}_{FK}|$. If $\gamma = 0$, no smoothing is needed; as $\gamma$ increases, more smoothing is needed.
Figure~\ref{Figure:smoothing} presents the results. 

The plots confirm our intuition: the $\textbf{X}_R$-based smoothing yields much lower test errors for both \textit{NoJoin} and \textit{JoinAll}--in 
fact, errors comparable to \textit{NoFK} and the Bayes error--for lower values of $\gamma$ ($<0.5$). But as $\gamma$ gets closer to $1$, the 
errors of $\textbf{X}_R$-based smoothing also increase but not as much as random hashing. Overall, these results suggest that even if foreign features 
are available, rather for using them directly for learning the model, we could use them as side information for smoothing $FK$ features. Overall,
these results suggest that it is possible to get ``the best of both worlds'': the runtime and usability gains of \textit{NoJoin} (as against \textit{JoinAll}, 
which unnecessarily also learns over the foreign features) along with exploiting the extra information provided by foreign features (if they are available)
for smoothing foreign key features.

\section{Related Work}


\paragraph*{Database Dependencies and ML}
The scenario of learning over joins of multiple tables without materializing the output of the join was studied in~\cite{orion,olteanuf,rendle,santoku},
but their goal was primarily to reduce runtimes of some ML techniques without affecting accuracy. It was also studied in~\cite{crossmine} but their
focus was on devising a new ML algorithm. In contrast, our work focuses on the more fundamental question of whether KFK joins can be avoided safely for 
existing popular ML algorithms. We first demonstrated the feasibility of avoiding joins safely in~\cite{hamlet} for linear models. In this work, we revisit that 
idea for higher capacity models and find that they are counter-intuitively \textit{more} robust than linear models to avoiding joins, not \textit{less} as the 
VC dimension-based analysis in~\cite{hamlet} suggested. We also empirically verify mechanisms to make foreign key features more practical.
Embedded multi-valued dependencies (EMVDs) are database dependencies that are more general than functional dependencies~\cite{dbtheorybook}. 
The implication of EMVDs for probabilistic conditional independence in Bayesian networks was originally described by~\cite{pearl} and further explored by~\cite{wong}.
However, their use of EMVDs still requires computations over all features in the data instance. In contrast, avoiding joins safely omits entire sets 
of features for complex ML models \textit{without performing any computations} on the foreign features.
There is a large body of work on statistical relational learning (SRL) to handle joins that cause duplicates in the fact table~\cite{srlbook}. But as mentioned before, 
our work focuses on the regular IID setting for which SRL might be an overkill.

\paragraph*{Feature Selection}
The data mining and ML communities have long worked on feature selection methods to improve ML accuracy~\cite{guyonbook,hastie}.
In contrast, our goal is \textit{not} to design new feature selection methods nor is it compare existing methods. 
Rather, we want to understand if KFKDs/FDs in the schema can enable us to avoid entire sets of features a priori for some popular complex classifiers.
This is a way of ``short-circuiting'' the feature selection process using database schema information to reduce the burden of data sourcing.
The trade-off between feature redundancy and relevancy is well-studied~\cite{guyonbook,leiyu,daphnekoller}. The conventional wisdom is that even a feature that is 
redundant might be highly relevant and hence, unavoidable in the mix~\cite{guyonbook}. Our work establishes, perhaps surprisingly, that this is \textit{not} the case 
for foreign features; even if a foreign feature is highly relevant, it can be safely discarded in most practical cases for decision trees, RBF-SVMs, and ANNs.
There is prior work on exploiting FDs in feature selection.
\cite{approxfds} infers approximate FDs using the dataset instance and exploits them during feature selection, FOCUS~\cite{focus} is an 
approach to bias the input and reduce the number of features by performing some computations over those features, while~\cite{consistencyfs} proposes
a measure called consistency to aid in feature subset search.
The idea of avoiding joins safely is orthogonal to these algorithms because they all still require computations over all features, while avoiding a join safely 
\textit{omits dependent features without even looking at them} and obviously, without performing any computations on them!
To the best of our knowledge, no feature selection method exhibits such a dramatic capability.
Scores such as Gini and information gain are known to be biased towards large-domain features in decision tree learning~\cite{dtreebias1} and different approaches 
have explored alternatives to solve that issue~\cite{dtreebias2}. Our problem is orthogonal because we study on how KFKDs/FDs enable us to ignore foreign features 
a priori safely. Even with the gain ratio score that is known to mitigate the bias towards large-domain features, our main findings stand.
Unsupervised dimensionality reduction methods such as random hashing or PCA are also popular~\cite{hastie,mitchellbook}. Our lossy compression techniques to 
reduce the domains of foreign key features for decision trees are inspired by such methods.

\paragraph*{Data Integration}
Integrating data and features from different sources for ML and data mining algorithms often requires applying and adapting techniques from
the data integration literature~\cite{diforml,dibook}. These include integrating features from different data types in recommendation 
systems~\cite{multitypefusion}, sensor fusion~\cite{sensorfusion}, dimensionality reduction during feature fusion~\cite{dimredfusion},
and techniques to control data quality during data fusion~\cite{lunabdi}.
Avoiding joins safely can be seen as one schema-based mechanism to reduce the integration burden by predicting a priori if a source table is unlikely 
to improve ML accuracy. It is a major open challenge to devise similar mechanisms can be devised for other types of data sources, say, using other 
forms of schema constraints, ontology information, and sampling. There is also a growing interest in making data discovery and other forms of 
metadata management easier~\cite{discovery, ground}. Our work can be seen as a mechanism to verify the potential 
utility of some of the discovered data sources using their metadata. We hope our work spurs more research in this direction of exploiting ideas from 
data integration and data discovery to reduce the data sourcing burden for ML tasks.



\section{Conclusions and Future Work}
We think it is high time for the data management community to look beyond just building faster 
or scalable ML systems and help reduce the pains of data sourcing and feature engineering for ML. 
Understanding how fundamental properties of data sources, especially schema information, affect 
ML behavior is one promising step in this direction. 
While the idea of avoiding joins safely has been adopted  in practice for 
linear classifiers, in this comprehensive experimental study, we show that it works even better for 
popular high-capacity classifiers, which goes against the intuition that high-capacity classifiers 
are typically more prone to overfitting. We hope that our results and analysis spur more discussions 
and new research on simplifying data sourcing for ML-based analytics.

As for future work, we plan to formally analyze the effects of KFKDs/FDs on high-capacity classifiers 
from a learning theoretic perspective. Other interesting avenues for future work include understanding 
the effects of more general database dependencies on classifiers, the effects of all database dependencies 
on regression and clustering models, and designing an automated ``advisor" for data sourcing for ML 
tasks, especially when there are heterogeneous data types and sources.

\pagebreak

\bibliographystyle{abbrv}
\balance
\bibliography{HamletVLDB}

\end{document}